\pdfoutput=1
\documentclass{aa}

\usepackage{color}
\usepackage{graphicx}
\usepackage{txfonts}
\usepackage{longtable} 
\usepackage{lscape}
 
\definecolor{Mygreen}{rgb}{0.75, 0.0, 0.0}
\definecolor{Mypink}{rgb}{1.0, 0.0, 0.5}
\definecolor{Myred}{rgb}{0.7, 0.0, 0.0}

\usepackage[breaklinks, citecolor=blue, linkcolor=Myred, urlcolor=Myred, colorlinks=true]{hyperref}

\bibpunct[; ]{(}{)}{,}{a}{}{;}
\defcitealias{pierre_xxl_2015}{XXL~Paper~I}
\defcitealias{pacaud_xxl_2015}{XXL~Paper~II}
\defcitealias{giles_xxl_2015}{XXL~Paper~III}
\defcitealias{lieu_xxl_2015}{XXL~Paper~IV}
\defcitealias{fotopoulou_xxl_2016}{XXL~Paper~VI}
\defcitealias{ziparo_xxl_2015}{XXL~Paper~X}
\defcitealias{eckert_xxl_2015}{XXL~Paper~XIII}
\defcitealias{lavoie_xxl_2016}{XXL~Paper~XV}
\defcitealias{adami_2017_submm}{XXL~Paper~XX}
\defcitealias{guglielmo_xxl_2017}{XXL~Paper~XXII}

\begin{document}

   \title{The XXL Survey \thanks{Based on observations obtained with XMM-Newton, an ESA science mission with instruments and contributions directly funded by ESA Member States and NASA.}}
\titlerunning{The XXL Survey, XXVIII}

   \subtitle{XXVIII. Galaxy luminosity functions of the XXL-N clusters }
   \author{M. Ricci\inst{\ref{oca}}\thanks{Corresponding author: \href{mailto:marina.ricci@oca.eu}{marina.ricci@oca.eu}}, C. Benoist\inst{\ref{oca}}, S. Maurogordato\inst{\ref{oca}}, C. Adami\inst{\ref{lam}}, L. Chiappetti\inst{\ref{milan}}, F. Gastaldello\inst{\ref{milan}}, V. Guglielmo\inst{\ref{lam},\ref{padova}}, B. Poggianti\inst{\ref{padova}}, M. Sereno\inst{\ref{INAF},\ref{bologna}}, R. Adam\inst{\ref{oca},\ref{cefca}},  S. Arnouts\inst{\ref{lam}}, A. Cappi\inst{\ref{oca},\ref{INAF},\ref{bologna}}, E. Koulouridis \inst{\ref{cea},\ref{cea_PD}}, F. Pacaud\inst{\ref{bonn}}, M. Pierre\inst{\ref{cea},\ref{cea_PD}} and M. E. Ramos-Ceja\inst{\ref{bonn}}
          }
\authorrunning{M. Ricci et al}

   \institute{\label{oca}Université Côte d'Azur, Observatoire de la Côte d'Azur, CNRS, Laboratoire Lagrange, Bd de l'Observatoire, CS 34229, 06304 Nice cedex 4, France 
              \and \label{lam}
                LAM, OAMP, Université Aix-Marseille, CNRS, Pôle de l’ Etoile, Site de Château Gombert, 38 rue Frédéric Joliot-Curie, 13388, Marseille 13 Cedex, France
            \and \label{milan}
             INAF - IASF Milan, via Bassini 15, I-20133 Milano,  Italy
              \and \label{padova}
             INAF - Astronomical Observatory of Padova, Vicolo Osservatorio 5 - 35122 - Padova, Italy
           \and \label{INAF}
            INAF - Osservatorio di Astrofisica e Scienza dello Spazio di Bologna, via Piero Gobetti 93/3, I-40129 Bologna, Italia
             \and \label{bologna}
            Dipartimento di Fisica e Astronomia, Alma Mater Studiorum -- Universit\`a di Bologna, via Piero Gobetti 93/2, I-40129 Bologna, Italia
                \and \label{cefca}
         Centro de Estudios de F\'isica del Cosmos de Arag\'on (CEFCA), Plaza San Juan, 1, planta 2, E-44001, Teruel, Spain
              \and \label{cea}
               IRFU, CEA, Université Paris-Saclay, F-91191 Gif sur Yvette, France
               \and  \label{cea_PD}
                Université Paris Diderot, AIM, Sorbonne Paris Cité, CEA, CNRS, F-91191 Gif sur Yvette, France
             \and \label{bonn}
             Argelander Institut für Astronomie, Universität Bonn, Auf dem Huegel 71, DE-53121 Bonn, Germany }

   \date{Received March 9, 2018; accepted May 26, 2018}

%##################################################################
%##################################################################
%###########################TEXT###################################
%##################################################################
%##################################################################

%##################################################################
%##################################################################
%ABSTRACT
%##################################################################
%##################################################################

  \abstract
% context heading (optional)
{The luminosity function (LF) is a powerful statistical tool used  to describe galaxies and learn about their evolution. In particular, the LFs of galaxies inside clusters allow us to better understand how galaxies evolve in these dense environments.
Knowledge of the LFs of galaxies in clusters is also crucial for clusters studies in the optical and near-infrared (NIR) as they encode, along with their density profiles, most of their observational properties.
However, no consensus has been reached yet about the evolution of the cluster galaxy LF with halo mass and redshift.}
% aims heading (mandatory) 
{The main goal of this study is to investigate the LF of a sample of 142 X-ray selected clusters, with spectroscopic redshift confirmation and a well defined selection function, spanning a wide redshift and mass range, and to test the LF dependence on cluster global properties, in a  homogeneous and unbiased way.}
 % methods heading (mandatory)
{Our study is based on the Canada--France--Hawaii Telescope Legacy Survey (CFHTLS) photometric galaxy catalogue, associated with photometric redshifts. 
We constructed LFs inside a scaled radius using a selection in photometric redshift around the cluster spectroscopic redshift in order to reduce projection effects. The width of the photometric redshift selection was carefully determined to avoid biasing the LF and depended on both the cluster redshift and the galaxy magnitudes. 
The purity was then enhanced by applying a precise background subtraction.
In order to enhance the signal we constructed composite luminosity functions (CLF) by stacking the individual LFs.
We then studied the evolution of the galaxy luminosity distributions with redshift and richness, analysing separately the brightest cluster galaxy (BCG) and non-BCG members. We fitted the dependences of the CLFs and BCG distributions parameters with redshift and richness conjointly in order to distinguish between these two effects.}
% results heading (mandatory)
{We find that the usual photometric redshift selection methods can bias the LF estimate if the redshift and magnitude dependence of the photometric redshift quality is not taken into account. 
Our main findings concerning the evolution of the galaxy luminosity distribution with redshift and richness  are that, in the inner region of clusters and in the redshift-mass range we probe (about $0<z<1$ and $10^{13}$M$_\odot<M_{500}<5\cdot10^{14}$M$_\odot$), the bright part of the LF (BCG excluded) does not depend much on mass or redshift except for its amplitude, whereas the BCG luminosity increases both with redshift and richness, and its scatter decreases with redshift.
}
{}


   \maketitle

%##################################################################
%##################################################################
%Section : INTRODUCTION
%##################################################################
%##################################################################
\section{Introduction}
\label{intro}

The galaxy luminosity function (LF) and its evolution with redshift, galaxy type, or environment
is one of the main tools for  constraining models of galaxy formation and evolution.

Knowledge of the LFs of galaxies in clusters is also important in cosmology, particularly in view of the future optical or near-infrared (NIR) wide-field surveys (e.g. \textit{Euclid}, LSST). The galaxy LFs of clusters, along with their density profiles, encode most of the observational properties of galaxy clusters in the optical. The LF and its evolution is therefore a key parameter in cluster detection. 
Moreover, in order to derive cosmological constraints from cluster counts, a precise and well-calibrated cluster mass estimate, based on an observable, is required.
The main mass proxies in the optical are the cluster richness \citep[e.g.][]{rozo_improvement_2009,andreon_scaling_2010} and optical-NIR luminosity  \citep[e.g.][hereafter XXL~Paper~X]{lin_near-infrared_2003, mulroy_galaxy_2017, ziparo_xxl_2015}, and these proxies  often require the knowledge of the cluster's LFs, for example by counting galaxies brighter than a characteristic magnitude or by integrating the luminosity function.
Thus, the LF of cluster galaxies is also a critical property that simulations need to reproduce if they are later used to  characterise cluster finder algorithms or calibrate observables. 

In a pioneering work, based on the \cite{press_formation_1974} work on the mass function, \cite{schechter_analytic_1976} proposed an  analytic expression to characterise the galaxy luminosity function, consisting of the product of a power law by a decreasing exponential function. It  is fully characterised by three parameters: the characteristic magnitude $M^*$ corresponding to the `knee' of the function , the slope $\alpha$ of the power law dominating at faint luminosities, and the characteristic density $\phi^*$.
Extensive work has been devoted  in recent decades to evaluating galaxy luminosity functions in different environments, from field to clusters, in different redshift ranges, and with different selection for galaxies  (colours and types). 
This resulted in a better theoretical  modelling of galaxy and structure formation and evolution \citep[see e.g.][]{menci_binary_2002, mo_dependence_2004}. 

Evolution of the LF with redshift is of particular interest as it is directly linked to the formation history of galaxies. It has been shown to be connected both to environment and to galaxy types. However, one of the main difficulties in the LF determination from photometric surveys is the correct evaluation of the background contamination, which is more critical for faint galaxies. Many analyses focusing on early-type  galaxies used the red sequence (the locus formed by early-type galaxies in colour-magnitude plane) to optimise the LF determination.  Most of them indicate that the fraction of passive galaxies in clusters changes with redshift, with a deficiency in low luminosity red galaxies for high redshift clusters with respect to low redshift ones \citep{de_lucia_buildup_2004, de_lucia_build-up_2007, stott_increase_2007, gilbank_red-sequence_2008, lu_recent_2009, rudnick_rest-frame_2009},  while some others disagree on this point  \citep[e.g.][]{andreon_build-up_2006,andreon_history_2008, crawford_red-sequence_2009}.  This  effect suggests that a large fraction of high redshift, low mass galaxies are blue, and progressively migrate to the red sequence at lower redshift. 
 
Photometric redshifts, whose quality has highly improved in the last decade, have led to significant progress in the determination of the LF of the whole population in the optical rest-frame, and of the relative  behaviour of the early- and late-type galaxy components \citep{rudnick_rest-frame_2009, martinet_evolution_2015, sarron_evolution_2017}.  Great insight at redshift $z > 1 $ was provided by analysis in the NIR rest-frame, which traces well the stellar mass \citep{muzzin_evolution_2008,mancone_formation_2010}.

Concerning the bright end of the LF, various  analyses converge to the fact that the characteristic magnitude redshift evolution up to $z\sim1$ can be described by passive evolution of a population formed in a starburst at high redshift \citep{de_propris_k-band_1999, de_propris_rest-frame_2007, de_propris_deep_2013, lin_evolution_2006}. This has been confirmed up to higher redshifts by analyses in the NIR and IR \citep{strazzullo_near-infrared_2006, muzzin_evolution_2008, mancone_formation_2010, mancone_faint_2012}. This last analysis also showed a flat faint end slope ( $\alpha \sim -1$ ) with no significant redshift evolution and stressed that the  evolution of $\alpha$ and $M^*$ have to be considered jointly for any interpretation in terms of evolution, due to the strong degeneracy between these parameters. 

The dependence of the galaxy luminosity function on cluster mass has also been investigated via  observed mass proxies such as richness, velocity dispersion, or X-ray luminosities and temperatures. Here again, a full consensus has not yet been reached, with some studies showing differences in the LF in clusters with low/high mass proxies \citep{valotto_luminosity_1997, croton_2df_2005, hansen_measurement_2005}, while others show little or no difference \citep{de_propris_2df_2003, alshino_luminosity_2010, moretti_galaxy_2015, lan_galaxy_2015}. 

Large cluster samples in X-rays or in the optical have recently become  available, spanning wide redshift and cluster mass ranges.
However, the study of the LF evolution in these samples is challenging because they are hampered by selection effects, leading to a bias between cluster masses and redshifts.
So far the approaches that have been used  to distinguish between mass and redshift effects are either  splitting the clusters and studying the LF in redshift and mass bins, as in \cite{sarron_evolution_2017}, or  using hierarchical Bayesian method that simultaneously models redshift evolution and cluster mass dependence, as in \cite{zhang_galaxies_2017}.

In the end, a full consensus has not yet been reached for  the evolution of the cluster galaxies LF with halo mass and redshift. The difficulty in comparing the results of the various analyses comes from the differences in sample selection, redshift and mass range, radius considered, method used to select galaxies, and statistical analysis performed. 
This strongly motivates the determination of the LF for a statistical sample of clusters with a homogeneous selection and a firmly tested methodology, and taking into account the bias between cluster mass and redshift.
In this paper we present the  analysis of the optical LFs of a sample of 142 galaxy clusters, detected in the X-ray by the XXL Survey and having spectroscopically confirmed redshifts, using the Canada--France--Hawaii Telescope Legacy Survey (CFHTLS) photometric data. This unique combination of surveys allows us to span a wide range of redshifts and X-ray luminosities (and thus masses). It also enables us to study the LF without being biased by optical  detection method. As we aim to characterise the luminosity function of the whole galaxy population, we make use of the state-of-the art photometric redshifts provided in the CFHTLS T0007 release. 
For this purpose, we have developed a new method that optimises the LF estimate  from photometric redshifts using the extensive spectroscopic data provided in the XXL project for calibration.  

The structure of the paper is as follows: we describe the data in Section \ref{data}; we present the method used to construct and parametrise  the LFs in Sections \ref{method} and \ref{fitting_process};  we show our results on the luminosity distribution and its dependence on the cluster  parameters in Section \ref{results}; and study the systematic effects in Section \ref{syst}. Finally, Sections \ref{disc} and \ref{ccl} are for the discussions and conclusions.

Throughout this paper, all magnitudes are expressed in the AB system \citep{1974ApJS...27...21O}. We use an evolutionary model as reference for the redshift evolution of the characteristic apparent magnitude $m^*$. This model was computed with  {\sc{LePhare}} using the elliptical galaxy SED template {\sc{burst\_sc86\_zo.sed}} from the {\sc{PEGASE2}} library \citep{1997A&A...326..950F}, with a redshift of formation $z_f=3$. We normalised the model using $K^*$ values from \cite{lin_evolution_2006} corrected to the AB system. This leads to a magnitude of $M^*_R=-21.36$ at $z=0$ in the $r'$ band. 
We use the notation $log$ and $ln$ for the common and natural logarithm respectively. Throughout this work we have used the cosmological parameters $H_0=70$ km.s$^{-1}$Mpc$^{-1}$, $\Omega_m=0.3$ and $\Omega_{\Lambda}=0.7$.

%##################################################################
%##################################################################
%Section : DATA DESCRIPTION
%##################################################################
%##################################################################
\section{Data description}
\label{data}

%##################################################################
%##################################################################
        \subsection{Custer sample}

The XXL Survey \citep[][XXL~Paper~I]{pierre_xxl_2015} is a XMM-Newton project designed to provide a well-defined sample of galaxy clusters out to z>1, suitable for precision cosmology  \citep[see][]{pierre_precision_2011} and for the analysis of galaxy evolution and active galactic nuclei.
The area covered is about $50$ square degrees divided in two fields of $25$ deg$^2$ each: XXL-North (XXL-N) and XXL-South (XXL-S).
The sensitivity of XXL is about $10^{-15}$ erg s$^{-1}$ cm$^{-2}$ in the [0.5--2] keV band ($3\sigma$ flux limit for point sources). Both fields benefit from an almost full imaging coverage in the optical (CFHTLS and HSC in the north, and BCS and DES in the south), NIR and far-infrared (e.g. WIRCAM, VISTA, Herschel/SPIRE, Spitzer), and millimetric (SPT in the south field).
The XXL cluster selection function was derived following the methodology developed for the XMM-LSS pilot survey and extensively tested on numerical simulations \citep[see][]{pacaud_xmm_2006}. The source detection algorithm was tested by comparing observations to Monte Carlo simulations, allowing to define different samples of extended sources according to their distribution in the extension-extension likelihood plane: the C1 and C2 class \citep[see also][XXL~Paper~II]{pacaud_xxl_2015}. 

The XXL cluster sample corresponding to the second XXL data release, XXL-365-GC, is presented in \cite{adami_2017_submm} (hereafter, XXL~Paper~XX). It contains the complete subset of clusters for which the selection function is well determined plus all X-ray clusters which have been, to date, spectroscopically confirmed. 
In the present study, we used the list of all C1 and C2 clusters from XXL-365-GC overlapping with the W1 field of the CFHTLS (i.e. clusters from the XXL-N field having a declination $\delta<-3.7$) and for which we have spectroscopic redshift confirmation. This led to a sample of $142$ clusters from $z=0.03$ to $1.06$, among which $93$ are classified as C1 and $49$ as C2. 

The redshift confirmation was made in \citetalias{adami_2017_submm}, using as a criterion the presence of at least three concordant redshifts   or having the redshift of the BCG. 
Hence, all clusters considered in the present study can be considered bona fide clusters: the C1 clusters constitute a `complete sample' (in the cosmological sense), while the current C2 sample is `pure' but not yet complete.

Throughout the study, the term `cluster' refers to an extended X-ray source having undergone spectroscopic confirmation. However, some of them may remain undetected by optical cluster finders if they are too poor or if there is an offset between the gas and the galaxies. Also, no distinction is made between groups and clusters. Finally, in the case of multiple structures, each substructure or group is identified as an X-ray cluster.

%##################################################################
%##################################################################
        \subsection{Cluster parameters from scaling relations}
        \label{param}

%_________________________________________________________________________________________
        \begin{figure}
           \resizebox{\hsize}{!}{\includegraphics{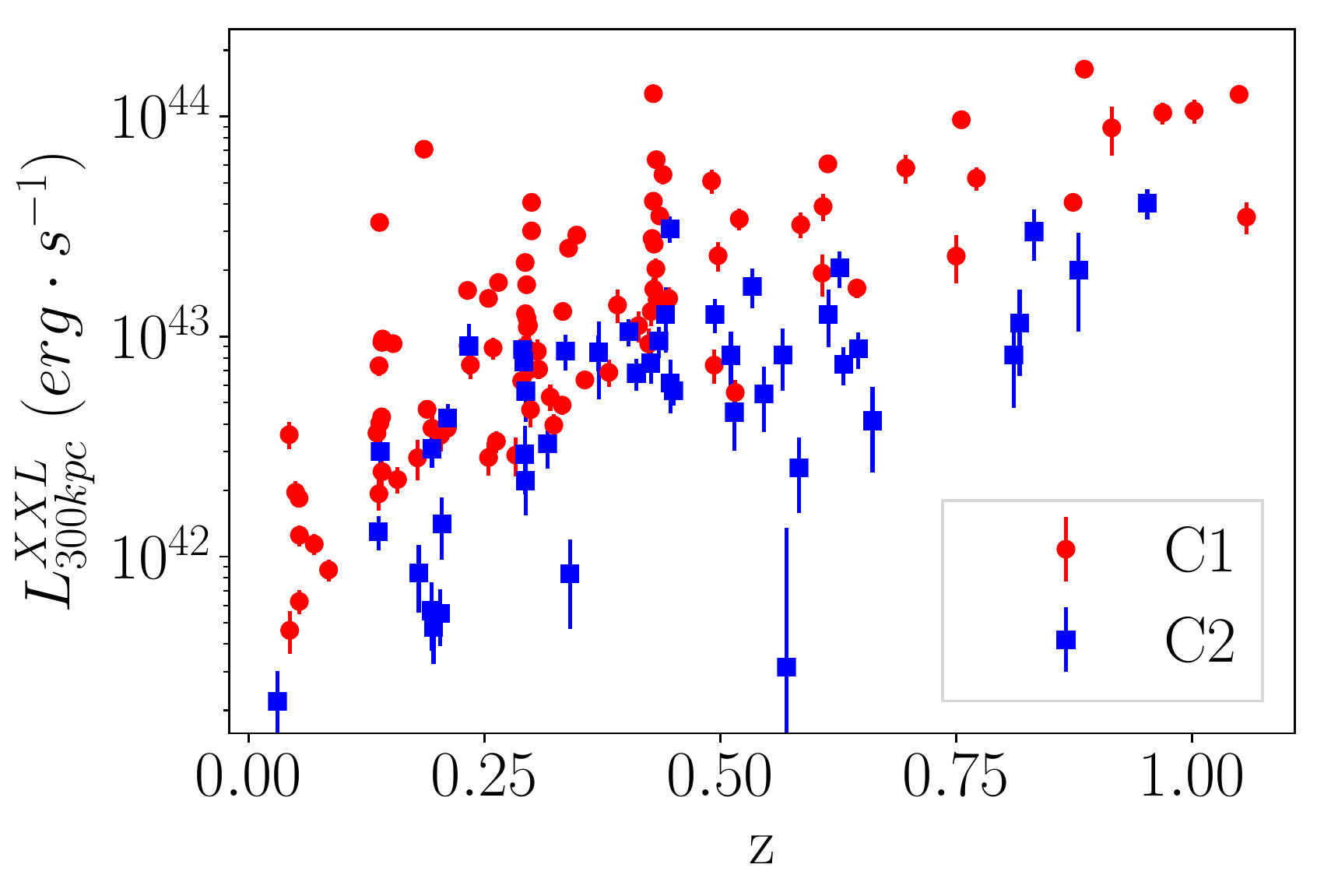}}
         \caption{\label{Lx_vs_z} X-ray luminosity in the $[0.5-2]$keV band computed in a $300$kpc aperture as a function of redshift for the cluster sample used in this study. Red points and blue squares represent clusters classified as C1 and C2, respectively (see text).}
        \end{figure}
        %script: local/XXL/catalogs/sample_plots.py
%_________________________________________________________________________________________

Because of  the faintness of some sources, it is not possible to obtain direct temperature estimates for all clusters. Therefore, in order to allow studies of the global properties of the full sample, 
we used cluster parameters extrapolated from an internally self-consistent system of scaling relations, based on the [0.5--2] keV X-ray count rates collected within a physical radius of $300$ kpc. The dedicated procedure is iterative and explained in detail in the Section 4.3 of \citetalias{adami_2017_submm}. In the following, these quantities are identified with the subscript ‘scal’.

Considering the good agreement between the parameters directly measured or extrapolated from scaling laws \citepalias[see figure 4 in][]{adami_2017_submm}, and  that we are interested in studying a global behaviour, we do not expect a major change in our results if we consider one or another type of measurements.

The parameter used in the rest of the study is $r_{500,scal}$\footnote{$r_{500}$ is defined as the radius of the sphere inside which the mean density is 500 times the critical density $\rho_c$ of the Universe at the cluster's redshift; $M_{500}$ is then by definition equal to $4/3\pi 500 \rho_c $ $r_{500}^3$.}, but we also mention the associated mass estimate, $M_{500,scal}$. For reference, Figure \ref{Lx_vs_z} shows the luminosities $L_{300kpc,scal}^{XXL}$  in the $[0.5-2]$ keV band and within $300$ kpc of our cluster sample, as a function of redshift. The red dots indicate the C1 clusters and the blue squares indicate the C2 clusters. It is important to note that since XXL is not a flux limited survey, but rather surface-brightness limited, the cluster locus in the $L_x-z$ plane does not follow a simple law \citep[see Figure 9 of][]{pacaud_xmm_2006}.

%##################################################################
%##################################################################
        \subsection{Galaxy catalogues}

%##################################################################
                \subsubsection{Photometric catalogue}

The optical counterpart of the XXL clusters comes from the CFHTLS, based on the optical and NIR wide-field imager MegaCam. The CFHTLS is composed of two surveys of different depth and area: the Deep Survey, split in four regions of $1$ deg$^2$ each, reaching an 80\% completeness limit in AB of $i’=25.4$ for point sources, and the Wide Survey, split in four regions of about $155$ deg$^2$ in total reaching an 80\% completeness limit in AB of $i’=24.8$ for point sources \citep[see][for more details]{hudelot_vizier_2012}. In this study data are taken from the W1 field of the Wide Survey, covering about $64$ deg$^2$ which overlaps most of the XXL-N survey.

The CFHTLS is conducted in five passbands: u*, g’, r’, i’, and z’, from approximately $300$
to $1000$ nm. The image stacking, calibration, and catalogue extraction was performed
by the Terapix data centre\footnote{\url{http://terapix.iap.fr/}}. We  used the latest version of the release, T007, which  provides better image quality and flux measurement precision than the previous releases, due to improved flat-fielding and photometric calibration techniques \citep[see][]{hudelot_vizier_2012}.
The source detection is made by {\sc{SExtractor}} \citep{1996A&AS..117..393B} on composite $g'r'i'$ images and the flux of the sources is then measured in each band using the same aperture. This technique provides reliable fluxes as the aperture is constant in each band, but may  lead to missing distant objects that appear only in the $z'$ band  \citep[see][]{szalay_simultaneous_1999}.

The masking of bright stars and image defects over the W1 CFHTLS field was performed in a semi-automatic way. Standard polygons, with a cross shape designed to enclose stellar spikes, were created for all stars brighter than $i'$=16. Polygon sizes are proportional to the star magnitude following an empirical relation validated by eye inspection. For the brightest stars and associated ghosts or for other types of defects (satellite trails, missing chips, field edges, etc.), polygons were designed by hand  to optimise the effective area to cross-match X-ray and optical data.
The final catalogue contains only unmasked objects  and the magnitude used is {\tt{MAG-AUTO}} which is a variable aperture Kron magnitude \citep[][]{kron_photometry_1980} and is well suited for galaxy studies.

%##################################################################
                \subsubsection{Photometric redshifts}

Precise  photometric redshifts taking advantage of multiwavelength photometry are available in the XXL framework \citep[see][XXL~Paper~VI]{fotopoulou_xxl_2016}. 
The quality of these photometric redshifts is optimised for the highest accuracy per galaxy; therefore, they are computed using a combination of wide and deep photometric observations (e.g. using the UKIDSS and VISTA surveys). They do not,  however,  cover the full CFHTLS W1 area homogeneously.
This strategy is not optimal for our statistical study which requires homogeneous redshift quality across the whole field.
We therefore used instead the photometric redshift catalogue associated with the CFHTLS W1 Survey, which is computed with five bands but presents a homogeneous quality across the field.

The estimation of the photometric redshifts in the CFHTLS W1 Survey was made using  {\sc{LePhare}} \citep[see][] {ilbert_accurate_2006,coupon_photometric_2009}.  {\sc{LePhare}} is a Fortran code that computes photometric redshifts using SED fitting. The procedure is done in two steps: first, theoretical magnitudes are computed according to the set of filters and the SED templates chosen;  second, theoretical magnitudes are fitted to the observed ones using a $\chi^2$ procedure, leading to a best fit SED template and a photometric redshift probability distribution function ($PDF_z$). 
An optimisation procedure, based on a spectroscopic training sample, is also performed to calibrate the SED template set, remove photometric systematic offset, and introduce priors on the redshift distribution.
A star/galaxy classification is provided by using only size criteria for bright objects and adding best fit SED criteria for fainter objects. Bad estimations lead to a contamination of about 1\% of stars in W1 and an incompleteness of galaxies of about 2.6\% \citep[see][]{coupon_photometric_2009}.

The set of SED templates used for the photometric redshift computations was constructed using  elliptical, spiral (SBc and Scd), and irregular galaxy templates from \cite{coleman_colors_1980} and a star-forming galaxy template from \cite{kinney_template_1996} \citep[\textsc{AVEROIN} \textsc{LePhare} SED package, as in][]{arnouts_swire-vvds-cfhtls_2007}. These six SEDs were then interpolated to produce a set of 62 templates.

The statistical choice to get discrete photometric redshift values from the $PDF_z$ was to take its median value $z_{PDF}$ instead of the mode of the distribution $z_{\chi^2}$  (as suggested in the T007 photometric redshift release explanatory document \footnote{\label{T007}\url{http://cesam.lam.fr/cfhtls-zphots/files/cfhtls_wide_T007_v1.2_Oct2012.pdf}}). Only the objects with photometric redshift computed with at least three photometric bands, a $\chi^2/dof$ value lower than 100 and a galactic type of SED were included in the final catalogue. This catalogue was then cut at a magnitude of $i'=24$.

%##################################################################
                \subsubsection{Spectroscopic redshifts catalogue}
%_________________________________________________________________________________________
        \begin{figure}
          \resizebox{\hsize}{!}{\includegraphics{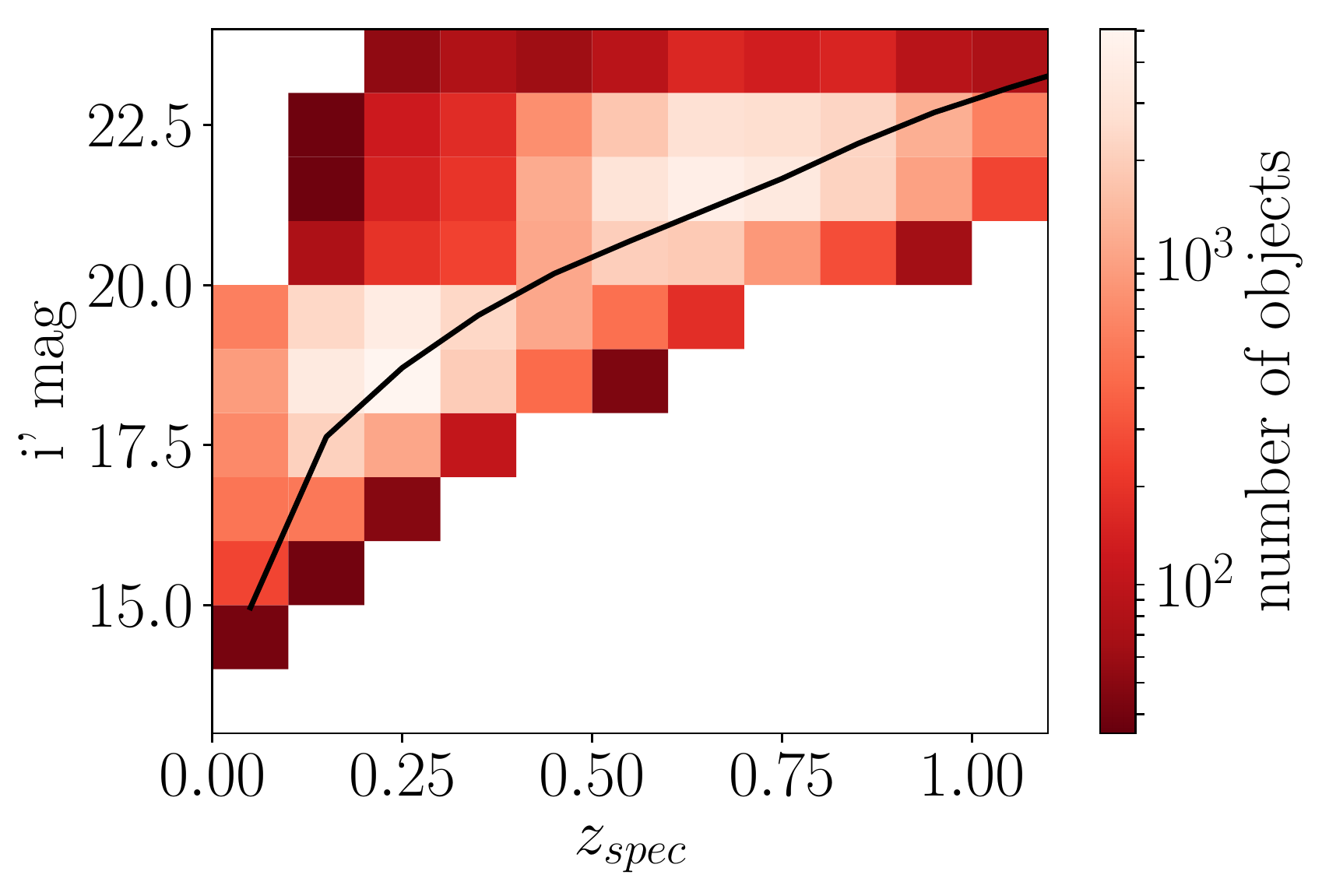}}
         \caption{\label{num_zspec}Number of objects with high quality spectroscopic measurements as a function of spectroscopic redshift and magnitude in the $i'$ band. The black line represents a fiducial evolution model for $m^*$.}
         %cosmos/stat_zphot_zspec_2d.py
        \end{figure}
%_________________________________________________________________________________________
The XXL spectroscopic data set used in this study is composed of several surveys and follow-ups conducted on the XXL-N field. It is described in detail in \citetalias{adami_2017_submm} and \citep[][hereafter XXL~Paper~XXII]{guglielmo_xxl_2017}, but a brief overview is given in the following.

 A large ESO programme has been allocated for XXL spectroscopic follow-up and cluster redshift confirmation. In addition to this programme, several dedicated projects have been conducted by XXL consortium members. The two major surveys available in the XXL-N field are the VIMOS Public Extragalactic Redshift Survey (VIPERS) and the AAOmega GAMA survey. They overlap respectively $16$ and $23.5$ square degrees of XXL-N. Other sources come mainly from VVDS Deep and the SDSS DR10 surveys. All these surveys are photometrically selected  and have different depths. VIPERS objects are selected using colour-colour diagrams to focus on galaxies between z = 0.5 and 1.2 with a limiting magnitude $I_{AB} = 22.5$. The other surveys have the following limiting magnitudes: $K_{AB} < 17.6$ \citep[see][]{baldry_galaxy_2010} for GAMA, $I_{AB} = 24.75$ for VVDS Deep and $g = 23$ for the SDSS-DR10 \citep[see][]{york_sloan_2000}. All the spectroscopic data were taken from the CESAM \footnote{\url{http://www.lam.fr/cesam/}} database.
 
Quality flags are available for the majority of surveys, albeit having different definitions \citepalias[see][for details]{guglielmo_xxl_2017}. No quality flags (zflags =-99) are available for the spectra coming from SDSS, Subaru, Alpha compilation, and NED.

%##################################################################
                \subsubsection{Spectro-photometric catalogue construction}
                \label{spectrophot}

The photometric and spectroscopic catalogues were matched according to their RA-Dec positions, allowing a maximum distance of one arcsecond. Multiple matches were treated by taking the nearest object. This procedure resulted in about 3\% of the photometric objects having a spectroscopic counterpart and a matched catalogue containing about $107500$ objects. 

The resulting spectro-photometric sample is highly dominated by GAMA at $z<0.5$ (28\% of the catalogue) and VIPERS at $z>0.5$ (57\% of the catalogue). Other contributions come from VVDS (at 4\%), SDSS (at 4\%), and 24 other origins (with less than 2\% of objects each).

We homogenised the spectroscopic quality flags in order to have equivalent quality definitions. In the following analysis, we discarded objects with quality flags corresponding to 5\% chances or more of having a false spectroscopic redshift, or without quality information. This high quality subsample includes 61\% of the objects from the spectro-photometric catalogue.

Figure \ref{num_zspec} shows the number of objects in the high quality subsample of the spectro-photometric catalogue as a function of redshift and magnitude in the $i'$ band. The black line represents an evolution model for the characteristic magnitude $m^*$ (see end of Section \ref{intro}).
 Figure \ref{bias} shows the comparison of photometric to spectroscopic redshifts for all galaxies from the high quality subsample.
 
%##################################################################
%##################################################################
%Section : CLUSTER GALAXY LUMINOSITY FUNCTION CONSTRUCTION
%##################################################################
%##################################################################
\section{Cluster galaxy luminosity function construction}
\label{method}

%##################################################################
%##################################################################
        \subsection{Luminosity function requirements}

The first critical step in the computation of cluster galaxies LFs is to properly count the right number of galaxies belonging to the cluster, in a given range of luminosity. In an ideal case, we would like to identify which galaxies belong to the cluster; however, precise cluster membership assignments are often difficult to perform, especially without spectroscopy. Alternatively, we can select highly probable cluster members, for example by using  photometric redshifts, and then statistically correct the field contamination by subtracting estimated counts from control background fields. The second critical step is to define the range of cluster galaxy luminosities  which  will not suffer from incompleteness. The methodology used to address these two points is developed in the following section.

%##################################################################
%##################################################################
        \subsection{Galaxy selection}

%##################################################################
                \subsubsection{Selecting galaxies using photometric redshifts}
                \label{photometric redshift_calib}
%_________________________________________________________________________________________
                \begin{figure}
                          \resizebox{\hsize}{!}{\includegraphics{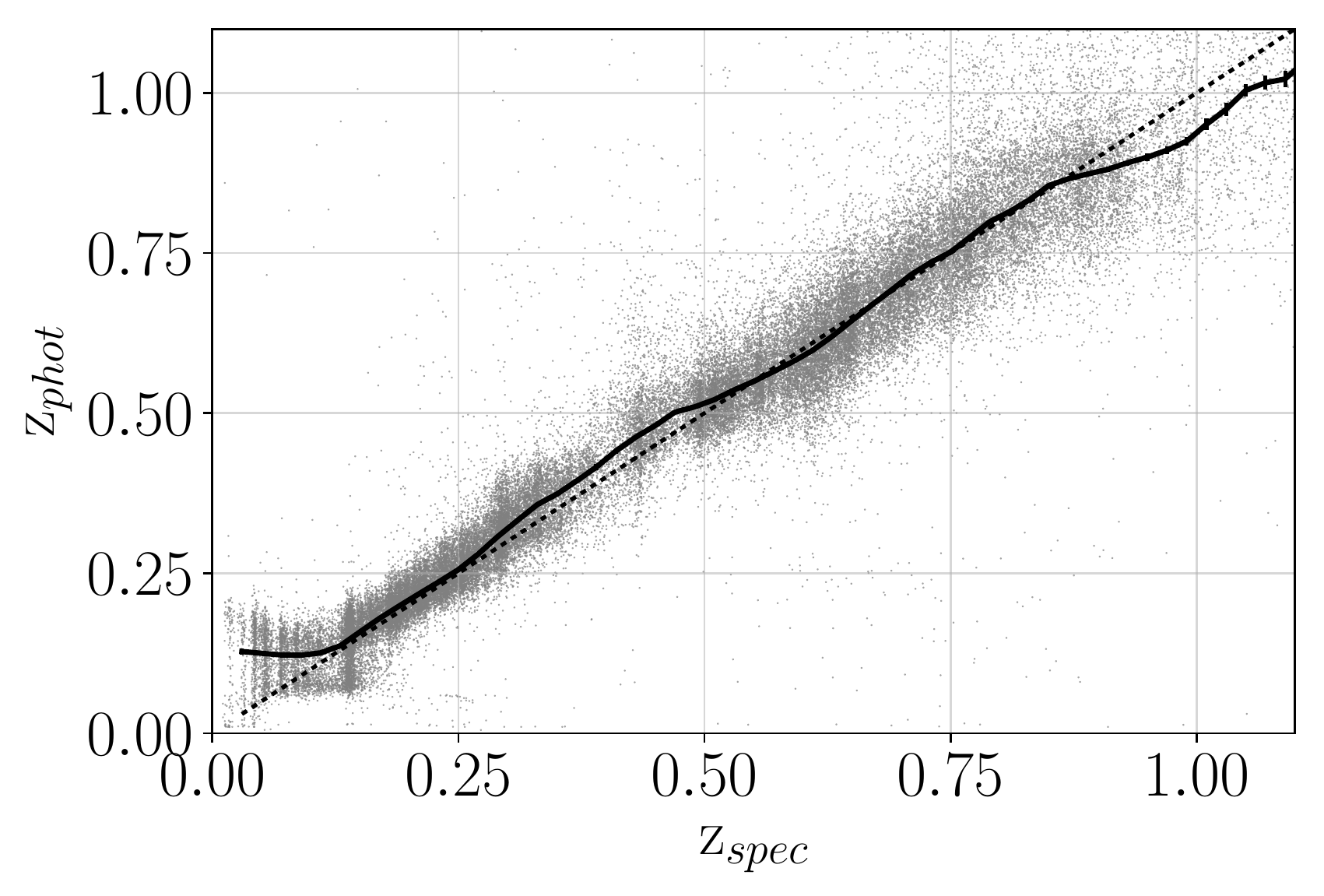}}
                 \caption{\label{bias}Relation between the photometric and spectroscopic redshifts, including all objects with a secure spectroscopic measurement (see text). Each grey  point represents a galaxy; the black line indicates the bias $b(z_{spec})$ and its error (only distinguishable at high redshift), which was computed assuming a normal distribution. The dot-dashed black lines indicates  $z_{phot}=z_{spec}$ for visualisation purposes.}
                %comos/stat_zphot_zspec_1d.py
                \end{figure}
%_________________________________________________________________________________________

%_________________________________________________________________________________________
        \begin{figure*}
        \begin{center}
          \includegraphics[width=140mm] {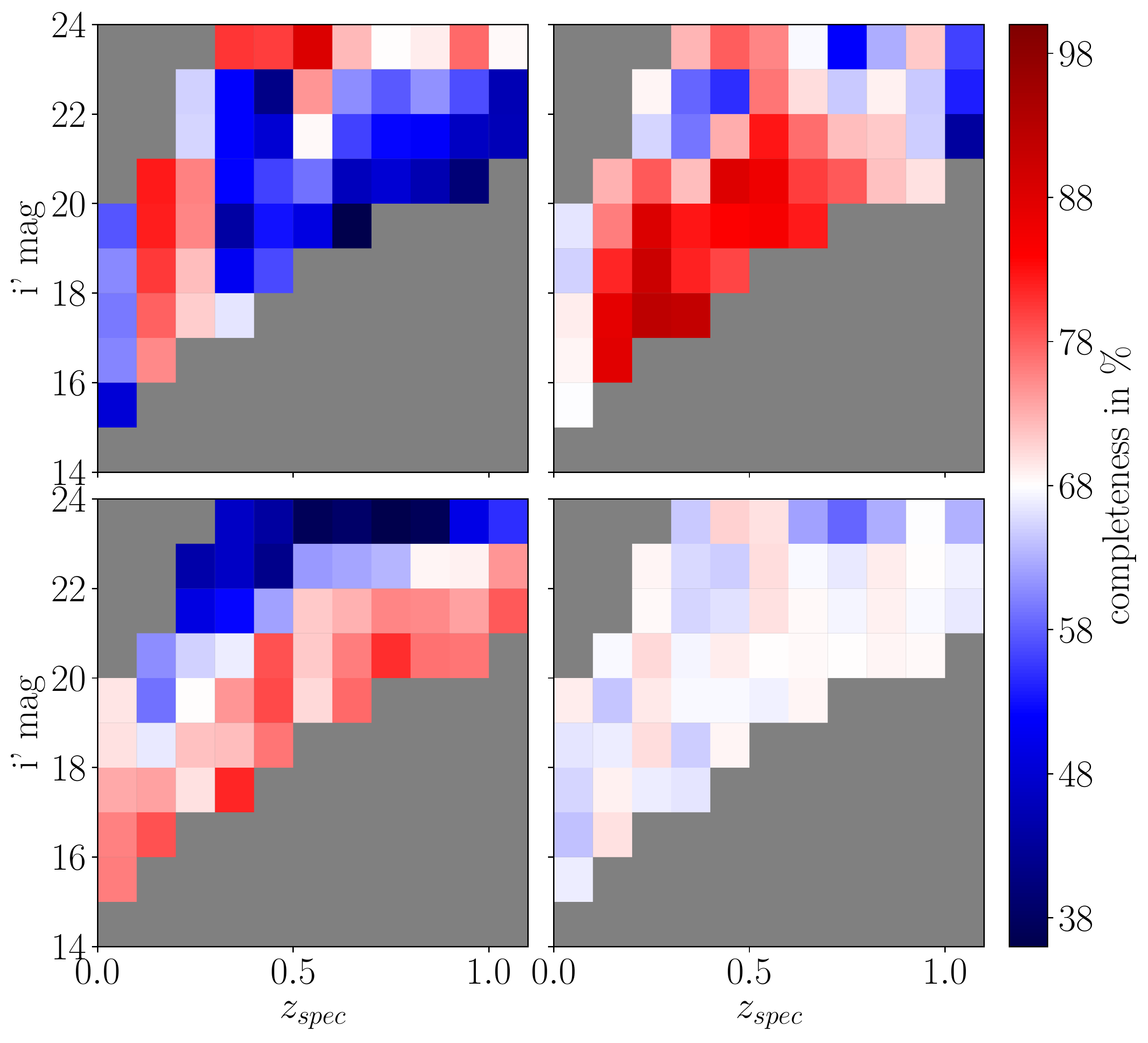}
         \caption{\label{compl1} Completeness (fraction of objects for which the photometric redshift is inside a given slice around the true redshift) of different galaxy selection methods as a function of spectroscopic redshift and magnitude in the $i'$ band, for selections at a $1\sigma$ (68\%) level. In red/blue the selection methods lead to over-/underestimate the number of objects. From top to bottom and left to right, the objects are selected using the PDZ errors (\emph{ZPDF} method), constant dispersions corresponding to $\sigma_{1/(1+z)}=0.04$ for $i'<22.5$ and $\sigma_{1/(1+z)}=0.08$ for $i'>22.5$ (\emph{cte} method), a dispersion computed as a $z_{spec}$ function (\emph{zfct} method), and a dispersion computed as a $(z_{spec},i' mag)$ function (\emph{zmfct} method, used in the rest of the study, see Figure \ref{disp}). The completeness is computed if there are at least 50 objects in the cell. }
         \end{center}
        \end{figure*}
        %comos/zphot_completness.py
%_________________________________________________________________________________________

As the number of available spectroscopic redshifts  differs greatly from cluster to cluster, we chose to use only photometric information to select member galaxies in order to keep a homogeneous selection. We also chose to select photometric redshifts based on discrete values within a range around the cluster spectroscopic redshift. A similar treatment was then applied to control
background fields. We discuss here various ways that have been used in other studies to define the photometric redshift range that assures a given level of cluster membership completeness, and we present our choice given our current data set. 

To select galaxies likely to be at the cluster redshift, we need to build the distribution $P( z_{phot,gal} | z_{clus})$, where $z_{phot,gal}$ are the galaxy photometric redshifts and $z_{clus}$ is the known spectroscopic redshift of the cluster. In the most general case this distribution depends on the galaxy magnitude, type, and redshift \citep[e.g.][]{ilbert_accurate_2006}. However, due to the large amount of spectroscopic data required to constrain these dependencies, the distribution is often averaged over magnitudes and types and modelled as a Gaussian distribution with a standard deviation given as $\sigma_{z} = \sigma_0 (1+z)$. If such a parametrisation is useful to describe the global performances of a photometric redshift algorithm, it may lead to inconsistencies in more detailed selections based on photometric redshifts.

It has been shown, for instance, that the fraction of catastrophic failures \citep[objects with $|z_{phot}-z_{spec}|>0.15(1+z_{spec})$, following the definition of ][]{ilbert_accurate_2006} and the dispersion both increase strongly with magnitude and redshift and get worse for galaxies with starburst SEDs \citep[see e.g.][]{ilbert_accurate_2006}. Moreover, the $P( z_{phot,gal} | z_{clus})$ distributions often show heavier tails than Gaussian distributions, which could lead to an additional source of incompleteness if not taken into account. 

Thanks to the XXL project,  we now have  a large associated spectroscopic catalogue that spans a wide range of redshifts, galaxy types, colours, and magnitudes, which we used to investigate the magnitude and redshift dependencies of the photometric redshift statistics. We used the spectrophotometric catalogue described in Section \ref{spectrophot}, selecting only secure spectroscopic redshifts. In all of the following analyses, the error on spectroscopic redshifts were considered  negligible with respect to that on photometric redshifts.

A first approach  to select photometric redshifts likely to be at a given spectroscopic redshift $z_{spec}$ (hereafter known as the \emph{ZPDF} method) is based on individual photometric redshift probability distribution functions ($PDF_z$) provided for each object in the CFHTLS T0007 release. The lower and upper photometric redshift estimation values $z_{p-}$ and $z_{p+}$ given in the catalogue are computed to enclose 68\% of the area around the median value ($z_{PDF}$). Therefore, 68\% of the galaxies at a given spectroscopic redshift $z_{spec}$ should verify:
        \begin{equation}
         z_{p-}<z_{spec}<z_{p+}
         \label{eq_z1} 
         \end{equation}
Based on the ($z_{phot} - z_{spec}$) statistics, we investigated three other ways to perform the photometric redshift selection, given the cluster spectroscopic redshift: {\it (i)} assuming the common Gaussian modelling with $\sigma_{z} = \sigma_0 (1+z)$ (hereafter known as  the \emph{cte} method); {\it (ii)} assuming a Gaussian modelling with $\sigma_{z}(z)$ computed in consecutive spectroscopic redshift bins (hereafter known as the \emph{zfct} method); and {\it (iii)} computing the 68th percentiles in bins of redshifts and magnitudes (hereafter known as the \emph{zmfct} method).

We defined the completeness of a given method as the ratio of the number of selected galaxies to the total number of galaxies in a given (redshift, magnitude) bin. As the selections are at a $1\sigma$ level, the completeness should be consistent with 68.2\%. The completeness computed as a function of magnitude and redshift  are shown by the four maps in Figure \ref{compl1}.

In the case of the \emph{ZPDF} method (upper left panel), we can see that the selection leads to an inhomogeneous completeness without a clear trend  with redshift or magnitude. Except for some regions, the completeness is generally lower than $68\%$, showing that the confidence intervals coming from the $PDF_z$ are usually underestimated. The lack of homogeneity observed in the completeness may be  caused by a potential bias of the photometric redshifts with respect to the spectroscopic ones. In this case, using the 68\% confidence limits around the median of the $PDF_z$ would lead to a photometric redshift window systematically shifted with respect to the spectroscopic value. 

The advantage of the three other methods, which  are directly computed from the ($z_{phot} - z_{spec}$) statistics, is that it is easy to introduce a bias correction that appears to be non-negligible in the present data set. Indeed, Figure \ref{bias} clearly shows the presence of systematics in several redshift windows. 
In particular at redshifts lower than $\sim 0.1$, photometric redshifts are systematically overestimated, while the opposite trend occurs at redshifts higher than $\sim 0.9$. We quantify the bias $b(z_{spec})$ as the median of $(z_{phot}-z_{spec})$. By computing it in the $(z_{spec},i')$ plane we saw that the bias depends mainly on the redshift; we thus computed it as a function of $z_{spec}$ only in running bins of $\Delta z =0.04$ from z=0.01 to z=1.31. We found that excluding or not the outliers before computing the median did not change the bias estimate in a significant way. The bias estimate and its error are shown by the black line in Figure \ref{bias}.
The  resulting bias function was introduced in the three methods described above (\emph{cte, zfct}, and \emph{zmfct}).

The completeness map corresponding to the \emph{cte} method is shown in the upper left panel  of Figure \ref{compl1}. We used a constant dispersion $\sigma_0=0.04$ for $i'<22.5$ and $\sigma_0=0.08$ for $i'>22.5$ (as suggested in the T007 photometric redshift release explanatory document $^{\ref{T007}}$) and we included the bias. The corresponding selection is the following:
        \begin{equation}
        -\sigma_0 (1+z_{spec})<z_{phot}-z_{spec}-b(z_{spec})<\sigma_0 (1+z_{spec})
        \label{eq_z2} 
        \end{equation}
We can see that the completeness is still not uniform: it is higher than $68\%$ for bright objects at low redshift and lower elsewhere, in particular at high redshift. This may be due to the fact that the dispersion is not simply evolving as $(1+z)$ with redshift. As an example, we can see in Figure \ref{bias} that the dispersion increases with the redshift, but is also higher in the $z_{spec}<0.1$ region.

The completeness map corresponding to the \emph{zfct} method is shown in the lower left panel of Figure \ref{compl1}. We estimated the dispersion $\sigma(z_{spec})$ as a function of redshift, in running bins of $\Delta z =0.04$ from z=0.01 to z=1.31, by computing the normal median absolute deviation (NMAD) in each bin, as $\sigma(z_{spec})=1.48$ median$|z_{phot}-z_{spec}-b(z_{spec})|$ and thus assuming Gaussianity. The corresponding selection criteria is 
        \begin{equation}
        -\sigma(z_{spec})<z_{phot}-z_{spec}-b(z_{spec})<\sigma(z_{spec}) 
        \label{eq_z3} 
        \end{equation}
We can see that the completeness is still not homogeneous but biased towards bright objects at every redshift. This occurs  because the dispersion is accurate where the number of objects is higher, the completeness pattern thus follows the redshift evolution of the median magnitude of the spectroscopic sample. We therefore removed the incompleteness due to redshift evolution of the dispersion but not that due to magnitude variation. This method is still not satisfying for LF study because it may artificially flatten the faint end slope. 

\begin{figure}
          \resizebox{\hsize}{!}{\includegraphics{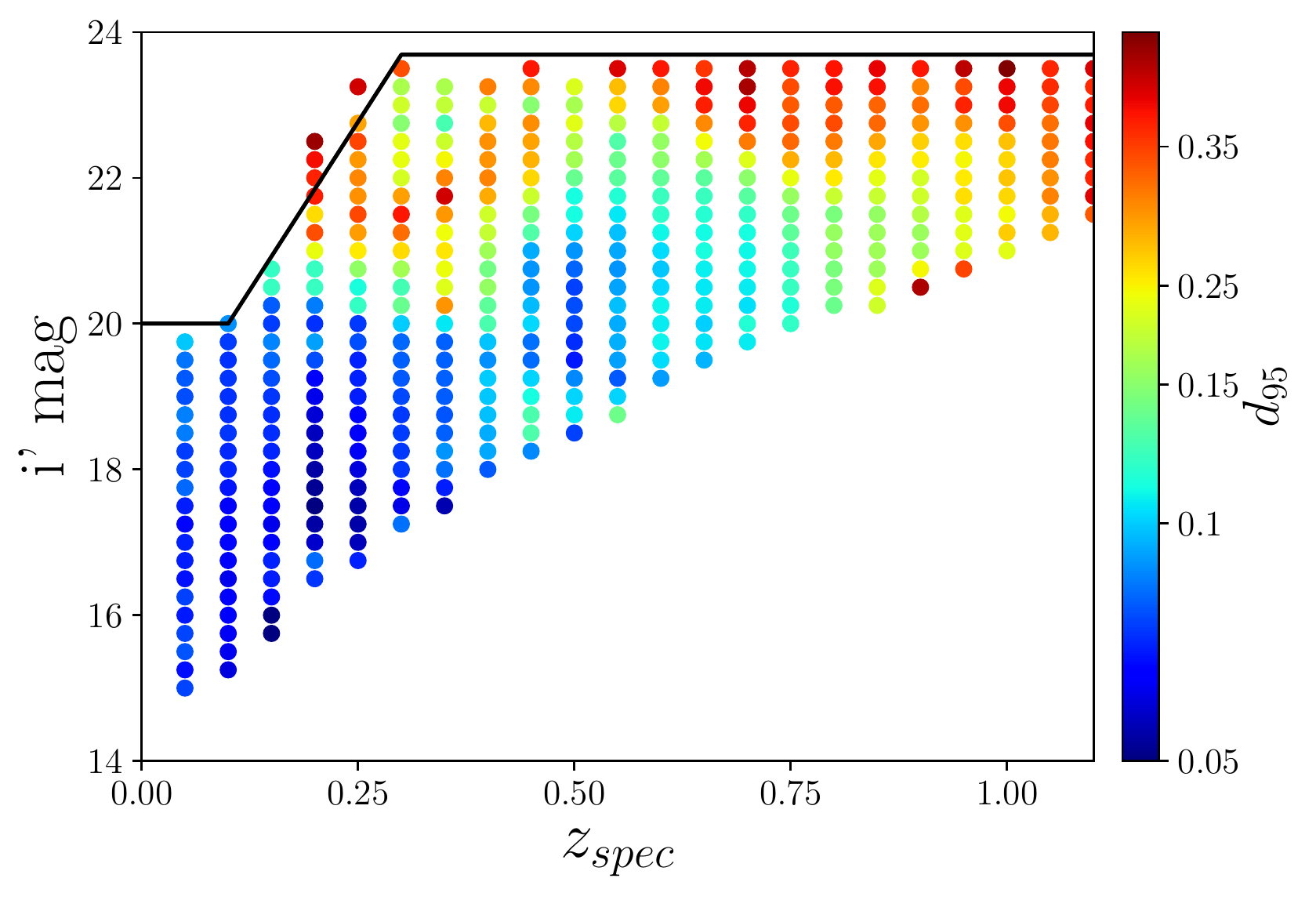}}
         \caption{\label{disp}Photometric redshift dispersion $d_{95}$, computed as the 95th percentile of
           $|z_{phot}-z_{spec}-bias(z_{spec})|$ in the $(z,i' mag)$ plane. The dots indicate the centres of the $\Delta (z,i' mag)=0.1\times0.5$ cells used to compute the dispersion (if they contain at least $30$ objects). The continuous black line shows the limiting magnitude we impose for the rest of the study.}
         % cosmos/stats_zphot_zspec_2d.py
        \end{figure}

Finally, the lower right panel of Figure \ref{compl1} shows the result of the \emph{zmfct} method. This method was designed to obtain the expected 68\% completeness map.  We computed the dispersion of the ($z_{phot}$ versus $z_{spec}$) distribution using percentiles instead of NMAD. This dispersion $d_n$ was defined as $d_n=P_n( |z_{phot}-z_{spec}-b(z_{spec})|)$, $P_n$ being the percentile of rank $n$. We computed it in the $(z_{spec},i')$ plane, using running cells of size $\Delta (z,mag)=0.1\times0.5$ if they contained at least $30$ objects. In order to limit the influence of catastrophic failures we filtered out the objects with dispersion values greater than 5 times the standard deviation of the $z_{phot}-z_{spec}-b(z_{spec})$ global distribution. We then interpolated the data to obtain a function of $(z_{spec},mag_i)$. The dispersion $d_{95}$, corresponding to 95\% completeness, is shown in Figure \ref{disp}; we note that the dispersion increases with redshift and magnitude independently. Unfortunately, we do not have enough spectroscopic data to constrain the dispersion for faint low redshift objects, as can be seen in Figures \ref{num_zspec} and \ref{disp}. Finally, we checked the completeness taking all objects for which 
        \begin{equation} -
        d_{68}(z_{spec},i')< z_{phot}-z_{spec}-bias(z_{spec})<
        d_{68}(z_{spec},i') 
        \label{eq_z4} 
        \end{equation} 
The resulting completeness (lower right panel of Figure \ref{compl1}) is indeed flat and compatible with 68.2\%. We therefore used this method to define the widths of the photometric slices of our raw membership assignments. We also investigated the effects on cluster luminosity functions of the three other selection methods in Section \ref{robust_method} and of the selection widths in Section \ref{robust_width}.

For each cluster with redshift $z_{clus}$ we thus selected possible member galaxies by taking all the objects satisfying
        \begin{equation}
        -d_{68}(z_{clus},i')<z_{phot}-z_{clus}-b(z_{clus})<d_{68}(z_{clus},i')
        \label{eq_zmfct} 
        ,\end{equation} 
where $d_{68}(z_{clus},i')$ is defined at the cluster redshift and changes according to the magnitude in the $i'$ band of each object considered. We computed dispersions corresponding to 95\%  completeness ($d_{95}$) in the same way as the 68\% complete ones. As we used percentiles and did not assume Gaussianity, $d_{95}$ is approximately but not simply equal to $2\times d_{68}$.
%_________________________________________________________________________________________
%_________________________________________________________________________________________

%##################################################################
                \subsubsection{Defining the background fields}
                \label{bck_sub}
%_________________________________________________________________________________________
        \begin{figure}
          \resizebox{\hsize}{!}{\includegraphics{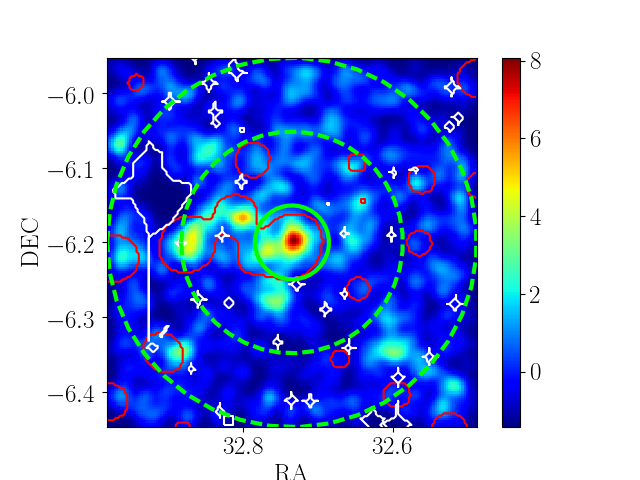}}
         \caption{\label{dens_map} Example of a density map of a cluster $10\times10$ Mpc$^2$ field, constructed using a Gaussian kernel of width $\sigma=0.1875 $ Mpc. The colourbar reflects the signal-to-noise ratio. Only galaxies with $m<m^*+3$ (or $L>0.06L^*$) are selected and the photometric redshift width depends on the galaxy magnitudes and is taken to ensure 68\% completeness. The red contours indicate the structures detected by  {\sc{WaZP}}, and the white ones show the masked regions. The green inner circle shows $1$ Mpc around the cluster X-ray centre and the dashed green lines delimitate the local background field from $3$ to $5$ Mpc. NB: We can see that this cluster is part of a superstructure.}
        \end{figure}
%_________________________________________________________________________________________

 In order to take into account the contamination of the cluster galaxy counts by foreground and background galaxies, we chose to statistically subtract background galaxy counts for each cluster.
The selection of local or global background fields to estimate the counts has been largely debated in the literature. Some differences may arise, on the one hand,  from
the fact that selecting a region too close to the cluster can bias the counts because of correlated signal from filaments or enlarged cluster outskirts, and on the other hand, because the clusters are embedded in the cosmic web and thus can lie on intrinsically high or low density regions compared to the whole field. \cite{goto_composite_2002} and \cite{popesso_rass-sdss_2005} showed that, in their rich cluster samples, the differences between the LF parameters obtained with the two methods were not significant. However, \cite{lan_galaxy_2015} found that their global background estimate, computed using  random fields of the same aperture size as their cluster fields, tended to underestimate the background level especially for low mass clusters.
In this study we thus chose to use local background fields enclosed in annuli of $3$ to $5$ Mpc around the cluster centres ($3$ Mpc $\sim 2.5 r_{500}$ for the more massive cluster in our sample). 

In some cases, the presence of groups in the periphery of the clusters may lead to an overestimation of the counts in the background fields. For this purpose, we adopted a similar treatment to that of \cite{de_filippis_luminosity_2011} and we ran the {\sc{WaZP}} cluster finder algorithm \citep[][]{Benoist_2017_in_prep} in  target mode on each cluster position and redshift, down to a magnitude of i = 24  to detect structures that may contaminate the background (see Figure \ref{dens_map} for an illustration). These structures were masked in the following analysis. 

 By masking, we do not take into account the possible projections  along the clusters' lines of sight and thus we may overestimate the galaxy counts in the cluster fields. However, the projected structures in cluster fields are less frequent than the structures in the background field; therefore, not removing the structures in the background will bias the counts low. \cite{castignani_new_2016} found that their membership assignment was less biased when removing the structures in the background, and \cite{rozo_redmapper_2015} found that, in their rich cluster sample, the correlated structures were contributing to approximatively 6\% of the clusters' richness. As we are working with relatively low mass clusters, for which projections are expected to be rarer, we thus expect less than 6\% contamination on our galaxy counts from possible correlated structures along the clusters' lines of sight, and we therefore neglected this effect. 
 
\begin{figure}
          \resizebox{\hsize}{!}{\includegraphics{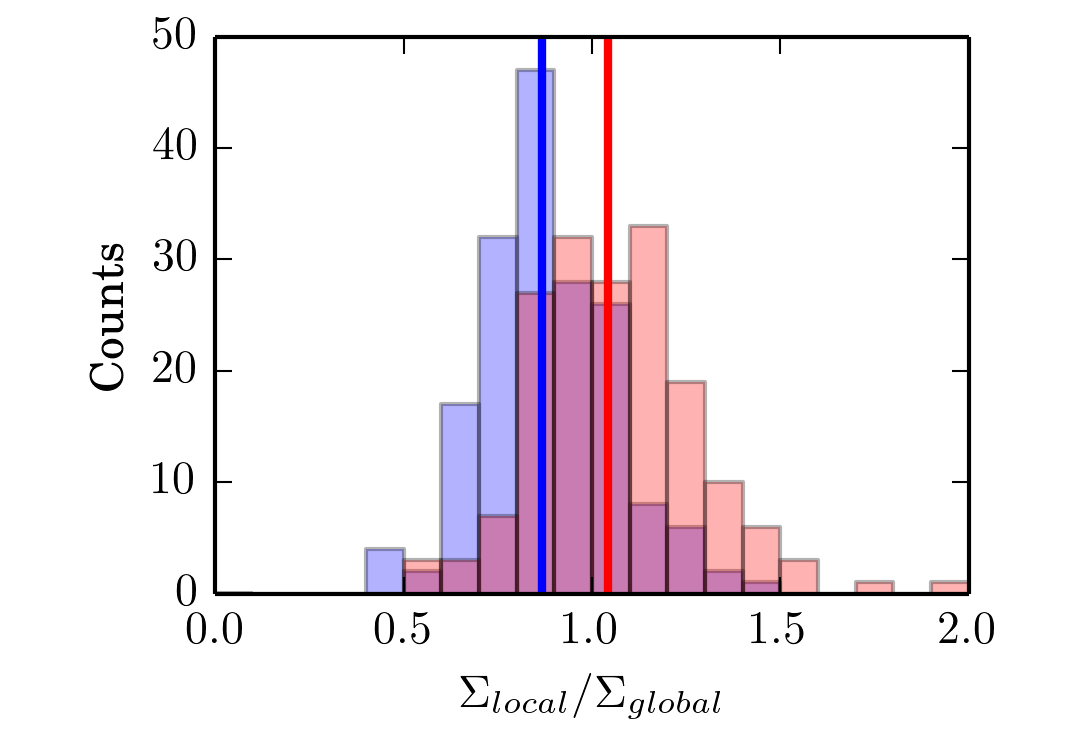}}
         \caption{\label{loc_glob} Histogram of the ratio between local and global background galaxy number densities. Global background refer to the whole CFHTLS W1 field, whereas local backgrounds refer to annuli of 3 to 5 Mpc centred on the X-ray cluster positions. The distribution of the ratio when structures are discarded from (taken into account in)  the local background fields is shown in blue (red). The solid lines indicate the median values of the ratios.}
        \end{figure}

For each cluster, we computed the effective local background area in Mpc$^2$, taking into account the photometric masks and the structure masks.
We compared counts in the local background fields to those obtained using the whole W1 field of $68$ deg$^2$, taking into account the photometric masks but not the structures. 
Figure \ref{loc_glob} shows the distribution of the ratio of local to global background galaxy densities when structures are discarded and taken into account from the local background fields (respectively in blue and red) and using galaxies brighter than $m^*$+1 ($0.4L_*$). As  can be seen, before removing the structures, the galaxy densities in the local background fields are in good agreement with the densities in the global field ($<\Sigma_{local}/\Sigma_{global}>\sim 1$). However, when the structures in the local fields are discarded,  galaxy densities become smaller than in the field on average . 
This is because we estimated the density in the global field as the mean density, which is sensitive to the presence of structures.
 The density ratio distributions are approximatively  log-normal and their widths denote the sample variance due to large-scale structures. We can see that some clusters are located in intrinsically underdense or overdense regions.

%_________________________________________________________________________________________
%_________________________________________________________________________________________

%##################################################################
%##################################################################
        \subsection{Defining the luminosity range}
        \label{sec_lum_range}

%##################################################################
                \subsubsection{Identification of the brightest cluster galaxy}
                
The luminosity of the BCGs has been shown to differ from the extrapolation of the LF of the other cluster members at high luminosity \citep[][]{schechter_analytic_1976} and many authors have chosen either  not to include them in the calculation of the LF or to treat them differently \citep[see e.g.][] {hansen_measurement_2005,wen_dependence_2014}. 
We therefore investigated the luminosity distribution of the BCGs separately and removed their contributions from the non-BCG members LFs. By definition, no cluster galaxy can be brighter, and thus we used the BCGs magnitudes as the bright limits of our luminosity ranges.

We identified the BCG for each cluster as the brightest galaxy in the apparent $i'$ band magnitude inside a projected radius of $400$ kpc from the X-ray centre, having either a spectroscopic redshift $z_{BCG}$, such as $z_{BCG}=z_{clus} \pm0.004\cdot(1+z)$ (with $z_{clus}$ the mean cluster redshift) or no spectroscopic redshift but a photometric redshift satisfying Eq. \ref{eq_zmfct}.
Visual inspection confirmed $134/142$ ($>94\%$) BCGs selected with these criteria and allowed us to identify the 8 others. 
%We present our sample BCG list in Appendix \ref{sec_tab_bcg}.

Our BCG list was compared to the one of \cite{lavoie_xxl_2016} (hereafter XXL~Paper~XV) as we have 40 clusters in common. We found different BCGs for $4/40$ ($10\%$) clusters. These discrepancies correspond to cases were several bright galaxies are present which makes the identification of the central galaxy difficult. 
The absolute magnitudes of the BCGs as a function of redshift are shown by the red points in Figure \ref{lum_range}.

%##################################################################
                \subsubsection{Limiting magnitudes}

The determination of the limiting magnitude is crucial for studies based on galaxy counts such as the luminosity functions. Photometric surveys are flux limited and if this effect is not taken into account, it can produce a spurious decline of the luminosity function at faint magnitudes.  We defined the completeness magnitude as the magnitude at which the completeness starts to decrease. In general completeness values are computed during the survey calibration phase. In the case of the W1 field, the completeness magnitudes at 80\% for extended sources, $mag_{80\%}$, are given  by the CHTLS-T0007 release explanatory document \citep[][]{hudelot_vizier_2012} and are $24.67 \pm 0.14,24.00 \pm 0.10$, and $23.69\pm 0.13$ in the $g'$, $r'$, and $i'$ band, respectively.

As we use photometric redshifts in this study, we have to take into account another source of incompleteness coming from the photometric redshift catalogue construction  because not all of the objects from the photometric catalogue have a good photometric redshift estimation (computed in three bands or more, with a $\chi^2/dof$ value lower than 100 and a galactic type of SED).  However, we find that this incompleteness is less than $3\%$ for every magnitude bin and we neglect it in our analysis.

As the low redshift/faint magnitude parameter space region is not well covered by spectroscopic surveys, the dispersion of the photometric redshifts in this region is not constrained, as can be seen from Figures \ref{num_zspec} and \ref{disp}.
 Therefore, we defined the limiting magnitude to be $m_{lim}=20$ at $z<0.1$, then linearly growing between $0.1<z<0.3$, up to $mag_{80\%}$ at $z>0.3$, as shown in Figure \ref{disp}. According to our fiducial evolution model for $m^*$, this cut allows us to include galaxies with $m>m^*+3$ (or $L<0.06L^*$) up to $z=0.6$.
 
\begin{figure}
            \resizebox{\hsize}{!}{\includegraphics{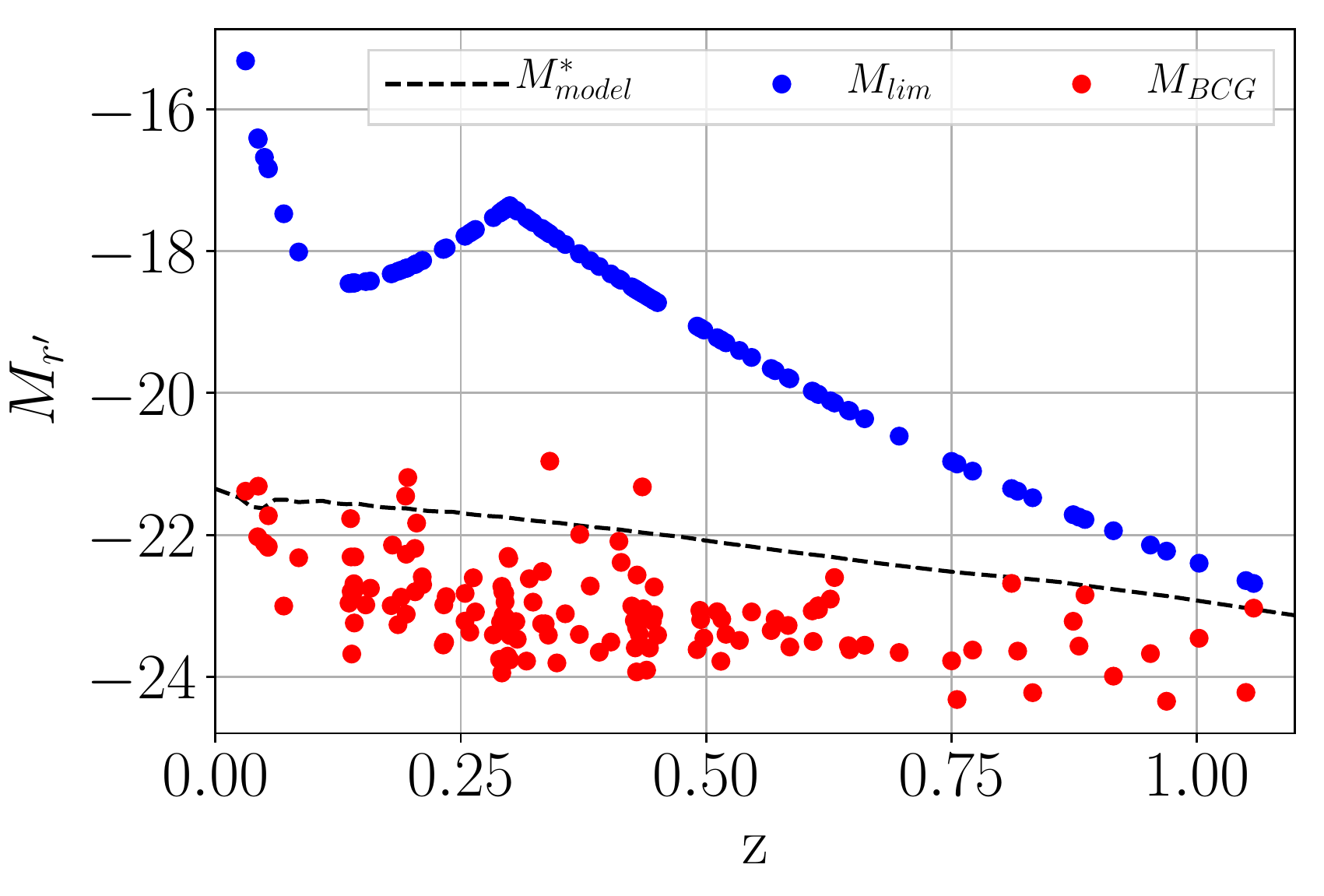}}
         \caption{\label{lum_range} Redshift evolution of the luminosity range in  which the LF are fitted. The red dots show the absolute magnitude of the BCGs of each cluster, whereas the blue ones indicate the limiting magnitude we imposed. A fiducial model (see end of Introduction) for the evolution of $M^*$ is indicated by the black dashed line for comparison.}
        \end{figure}

We converted the limiting magnitudes $m_{lim}(z)$ in absolute magnitudes following $M_{lim}(z)=m_{lim}(z)-\mu(z)-max(K_{corr})(z)$ 
with $\mu$ the distance modulus and $K_{corr}$ the k-correction. The model taken for the k-correction is the one used by  {\sc{LePhare}} to compute the absolute magnitudes and depends on galaxy type. To be conservative we took the maximum value of the $K_{corr}$ at each redshift, corresponding to that obtained for elliptical galaxies. The limiting absolute magnitude for each cluster, as a function of redshift, is shown by the blue points in Figure \ref{lum_range}. We can see that below $z=0.67$ the luminosity range is always wider than $\sim 3$ mag.

%_________________________________________________________________________________________
%_________________________________________________________________________________________

%##################################################################
%##################################################################
        \subsection{Counting galaxies}

%##################################################################
                \subsubsection{Galaxy counts in absolute magnitude}
                \label{abs_mag}
                
As  {\sc{LePhare}} uses SED modelling to compute absolute magnitudes, in order to have the absolute magnitude constrained by the observational data at $\lambda_{rest}$ we need to have $\lambda_u'<(1+z) \lambda_{rest}< \lambda_{z'}$, with $\lambda_u'$ and $\lambda_{z'}$ the wavelengths of the $u'$ and $z'$ filters and $z$ the redshift of the object considered. This condition is satisfied up to high redshift for the bluest bands. However, redder bands are known to be more representative of the stellar mass because they are less affected by star formation.
 This is why we chose to use the rest frame $r'$ band, which is constrained up to $z\sim 0.67$.
 
We assumed that each cluster member is at the mean redshift of the cluster.  We therefore used the value of the absolute magnitude provided by  {\sc{LePhare}} and computed with the photometric redshift estimation, and we corrected it by the redshift distance modulus offset. 

After selecting the potential member galaxies for each cluster using their photometric redshifts and the method described in Section \ref{photometric redshift_calib}, we statistically removed the contribution from the background galaxies.
To do this,  for each cluster field we defined the probability $P_{out}$(i' mag) of not being a cluster member as the galaxy number density ratio of the background to the cluster fields, as a function of apparent magnitude. The associated probability density functions were constructed using a Gaussian kernel density estimator with a standard deviation of $0.5$ mag.We then assigned  each potential member a random number $n$ between zero and one, and compared it to the probability $P_{out}$(i' mag) at the galaxy apparent magnitude. If $n<P_{out }$, the galaxy was discarded from the counts. This procedure was repeated 100 times: the counts were taken as the average values and their statistical error contribution were taken as the standard deviation. 

Finally, the counts were made inside projected $r_{500}$ radii and in absolute magnitude bins of $0.5$ mag. The number of galaxies per bin was normalised by the bin size and cluster area  to obtain the galaxy surface density $\phi$, expressed in $N_{gal}$~mag$^{-1}$Mpc$^{-2}$. The associated error in each bin $\Delta \phi_{j}$ was defined as the quadratic sum of the Poissonian and the statistical errors on the counts, normalised by the bin size and cluster area.

%##################################################################
                \subsubsection{Composite luminosity functions}
                \label{method_CLF}

In order to  investigate the dependence of the LF with cluster properties and enhance the signal-to-noise ratio, we chose to create composite cluster luminosity functions (CLFs). The stacking procedure was made using the method described in \cite{colless_dynamics_1989} in order to obtain CLFs extending up to the faintest magnitude limits of our sample, and thus use all available data, as recommended by \cite{popesso_rass-sdss_2005}: We define the following parameters:

\begin{itemize}

\item The galaxy surface density in the $j$th magnitude bin of the composite luminosity function 
         \begin{equation}
         \label{colless}
        \phi_{j}=\frac{\phi_{0}}{n_j} \sum_{i}\frac{\phi_{ij}}{\phi_{i0}} 
         ,\end{equation}
where $\phi_{ij}$ is the galaxy surface density in the $j$th magnitude bin of the $i$th cluster, $n_j$ is the number of clusters contributing to the $j$th magnitude bin, $\phi_{i0}$ is the normalisation of the $i$th cluster, and $\phi_{0}$ is  the mean normalisation $\phi_{0}= <\phi_{i0}>_{i}$  \citep[whereas in][$\phi_{0}=\sum_{i}\phi_{i0}$]{colless_dynamics_1989}. The normalisation $\phi_{i0}$ is defined as the sum of the galaxy surface densities in all the bins brighter than a limiting magnitude. This magnitude is tuned to be brighter than the limiting magnitudes of all the individual LFs in the stack.  Possible clusters for which $\phi_{i0}=0$ are not included in the CLF.\\

\item The statistical error associated with $\phi_{j}$  
          \begin{equation}
        \delta \phi_{j}=\frac{\phi_{0}}{n_j}\left[ \sum_{i}\left(\frac{\Delta \phi_{ij}}{\phi_{i0}}\right)^2\right]^{1/2}
         ,\end{equation}
 where $\Delta \phi_{ij} = \Delta \phi_{j}$ for the $ith$ cluster.
 
 \end{itemize}
 
Another source of errors comes from the intrinsic scatter between individual cluster LFs inside the CLF. To estimate this error we computed the CLF counts for 1000 resamplings of the stack using bootstrap. The final CLF counts were defined as the medians of the 1000 CLF realisation values, and the standard deviations $\sigma_{j}$ were used as the CLF intrinsic scatter per magnitude bin indicators.

The final errors in each magnitude bin of the CLF were taken as the quadratic sums of the statistical errors  and the intrinsic scatter, $\sqrt{\delta \phi_{j}^2+\sigma_{j}^2}$. In general, the statistical errors are dominant in the bright part of the CLFs and the intrinsic scatter is dominant in the faint part.

%##################################################################
                \subsubsection{Definition of the cluster richness}
                \label{dens}
                
In the following analysis, we  investigate the LF dependences on the  general properties of the clusters. For this purpose we chose to use the richness, which is a quantity naturally linked to the LF and a cluster mass indicator.
The richness is a very promising cluster mass proxy \citep[see e.g.][]{rozo_improvement_2009,andreon_richness-mass_2012} and has the advantage of being directly derived from the same photometric galaxy catalogue used for LF determination.
 
Precise membership assignment for our X-ray cluster sample is beyond the scope of this paper, but we instead wish to quantify the galaxy excess at the positions of extended X-ray source detections. 
Therefore, richness values $\lambda_{r}$ were computed using the differences  in galaxy density numbers between the cluster and background fields and their associated errors $\Delta\lambda_{r}$ were taken as Poissonian errors

$\lambda_{R}=\pi\cdot R^2\cdot\left(\Sigma_{cf}-\Sigma_{bf}\right)$ and $\Delta\lambda_{R}=\pi\cdot R^2\cdot\left(\frac{\Sigma_{cf}}{\sqrt{N_{cf}}}+\frac{\Sigma_{bf}}{\sqrt{N_{bf}}}\right)$~,

with $R$ the projected radius inside which the cluster field is defined, $\Sigma_{cf}$ and $\Sigma_{bf}$ the cluster and background field galaxy number densities, and $N_{cf}$ and $N_{bf}$ the cluster and background field galaxy number counts.

To compute richness values, we used the redshift and magnitude dependent photometric redshift dispersion presented in Section \ref{photometric redshift_calib} and we only selected galaxies with $m<m^*+1$ (or $L>0.4L^*$) in order to be complete up to $z\sim1$ and enhance the density contrast with respect to the field. Various aperture radii were explored, as we need to make a compromise between large radii that  introduce interlopers and noise and small ones that are sensitive to X-ray--optical centring offset. Finally, we chose to use a constant physical radius to have a mass proxy independent from scaling laws, with a size of $0.5$Mpc,  compared to the median $r_{500}$ of our sample ($\sim 0.6Mpc$). In the rest of the study, the richness is denoted by $\lambda_{0.5Mpc}$. 

%##################################################################
%##################################################################
%Section : LUMINOSITY FUNCTIONS (LFS) FITTING PROCEDURE
%##################################################################
%##################################################################
 \section{Luminosity function  fitting procedure}
\label{fitting_process}

%##################################################################
%##################################################################
        \subsection{Parametrisation by a Schechter function}

In order to characterise the CLFs and to compare them to other studies we parametrised them by Schechter functions \citep[][]{schechter_analytic_1976} 
        \begin{equation}
        \phi(L)dL=\phi^*\left(\frac{L}{L^*}\right)^\alpha exp\left(-\frac{L}{L^*}\right)\frac{dL}{L^*}
        ~\text{, }
        \end{equation}
and as $\frac{L}{L^*}=10^{0.4(M^*-M)}$ the function in terms of absolute magnitude can be expressed as
         \begin{equation} 
         \phi(M)dM=0.4 ln(10)\phi^*10^{0.4(M^*-M)(\alpha+1)}e^{-10^{0.4(M^*-M)}}dM 
        \end{equation}
with $\phi^*$ the characteristic number density, $M^*$($L^*$) the characteristic absolute magnitude (luminosity)  and $\alpha$ the faint end slope. 

Several authors, e.g. \cite{popesso_rass-sdss_2005}, have found that luminosity and stellar mass functions are best described by double Schechter functions in order to model separately the behaviour of their bright and faint parts. However, we do not reach sufficiently faint magnitudes to need this double parametrisation and consider a single Schechter component sufficient to describe our data.

 The contribution from the BCGs was removed and magnitude bins with less than 4.5 clusters contributing were not taken into account in the fit. Unless specified, the parameters $\phi^*$, $M^*$, and $\alpha$ were set free and constrained at the same time.

%##################################################################
%##################################################################
        \subsection{Computation of parameters probability density functions}

In order to properly define the errors on our parameters, we chose to estimate their probability density functions (PDFs). To do so, we computed $\chi^2$ values on $\phi^*-\alpha-M^*$ 3D grids. 
Due to the shape of our parameter likelihood, the values  are sensitive to the so-called volume effect;      depending on the statistical approach we used  to obtain the parameter's PDFs, we do not get the same results. In our case, as we use a grid that does not sample  the likelihood profiles finely enough, we used marginalisation to obtain the PDF of the parameters.
We thus marginalised over one parameter to compute the error ellipses around the other two, and marginalised over two parameters to obtain the PDF of the other one.

The sizes of the grids were chosen to encompass the 99\% likelihood contours, and we verify that if this criterion is satisfied the choice of the size does not affect the results. Also, the size of the cells has to be small enough so that the numerical errors can be neglected.

In the rest of the study we chose to use the median of the PDF as our statistical approach to get discrete values from the full likelihoods, as it is stable and not very sensitive to the grid sampling (we discuss the choice of statistical estimators in Sections \ref{robust_stats} and \ref{sys_discuss}). The reported errors on the parameters are then the 16th and 84th percentiles. The grids were chosen to contain $101\times101\times101$ points and to be bound by $\phi^*= [0,125]$, $\alpha = [-3.5 , 3.5],$ and $M_R^*=[-32 ,-18]$ when binning in redshift and $\phi^*= [0,35]$, $\alpha = [-1.75 , -0.25],$ and $M_R^*=[-32 ,-19]$ when binning in richness. Due to the low S/N and number of points of the CLF  in the highest redshift bin (see Section \ref{results}) the parameter likelihood was sampled only up to ~95\%.

%##################################################################
%##################################################################
        \subsection{Construction of parametrised composite luminosity functions}
The shapes of the parametrised composite cluster LFs were drawn by sampling the $\phi^*-\alpha-M^*$ space, according to the $\chi^2$ values. We computed 1000 realisations of the parameter set. We then used the median of the resulting LFs as the parametrised CLF profile, and we drew the 68\% confidence intervals around it using percentiles.

%##################################################################
%##################################################################
%Section : RESULTS AND ANALYSIS
%##################################################################
%##################################################################
\section{Composite luminosity functions and dependence on cluster parameters}
\label{results}

Composite luminosity functions were computed for the entire cluster sample with different selections. The methodology used is described in Sections \ref{method} and \ref{fitting_process}. Galaxies were selected using photometric redshift dispersion ensuring 95\% completeness ($d_{95\%}$, see Eq. \ref{eq_zmfct} and following text). The counts were made in projected $r_{500}$ in order to sample the same region for each cluster to avoid  mixing radial dependences with other effects (see e.g. \citealt{hansen_measurement_2005}, \citealt{popesso_rass-sdss_2006}, \citealt{Barkhouse_luminosity_2007}). We restricted the study to the clusters with redshift $z<0.67$ in order to have accurate estimations of the absolute magnitude in the rest frame $r'$ band (see Section \ref{abs_mag}) and treated the other clusters separately. 
In the following sections we analyse the composite luminosity function of the general sample ($z<0.67$) and 
investigate the dependence of the BCG and non-BCG luminosity distributions with both redshift and richness.

%##################################################################
%##################################################################
        \subsection{Composite luminosity function of the general sample}
        \label{full_sample}
%_________________________________________________________________________________________
                \begin{figure}[h]
                \begin{center}
                  \includegraphics[width=88mm]{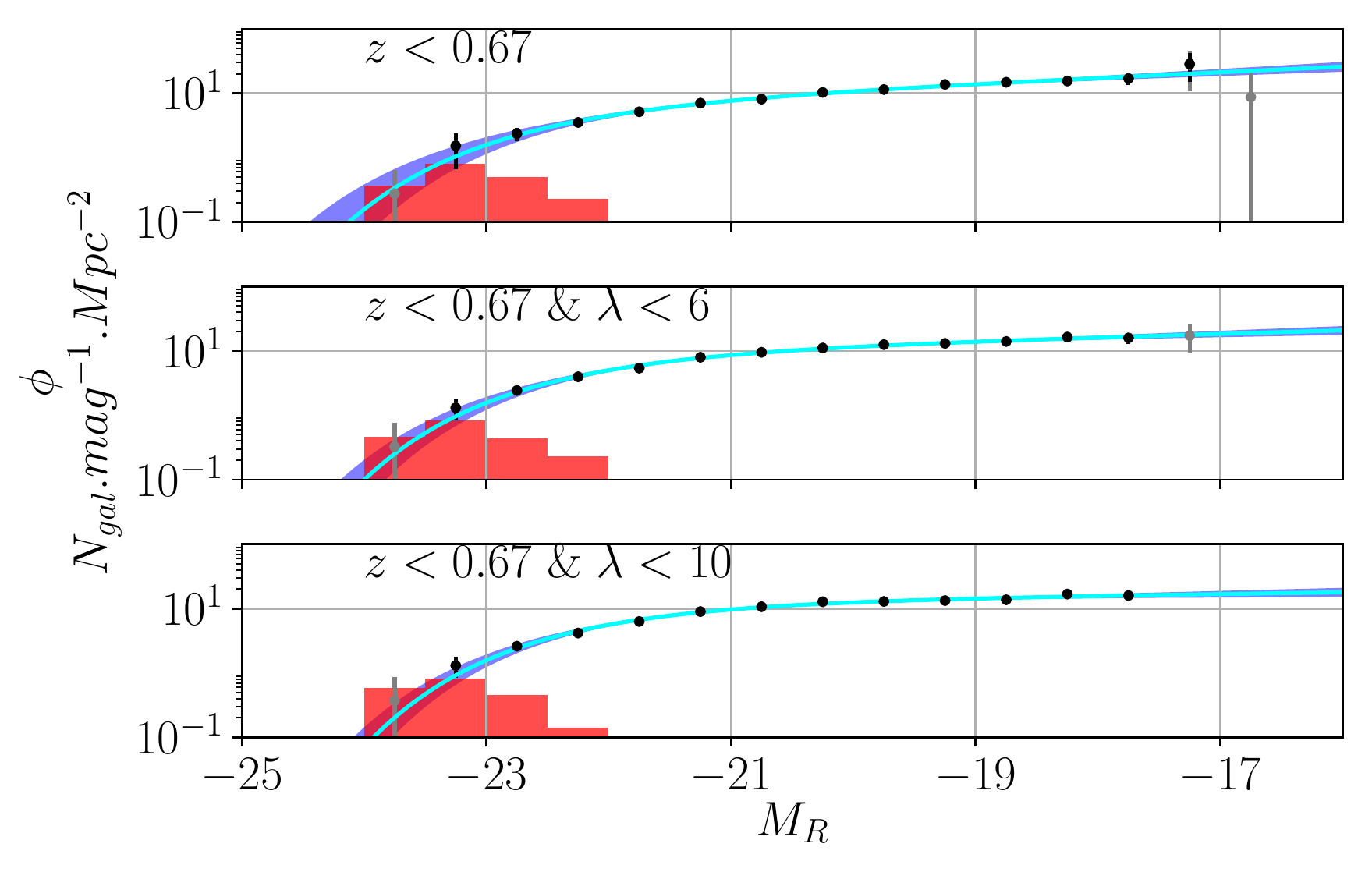}
                 \caption{\label{CLF_general} Composite luminosity functions including all clusters with a redshift lower than 0.67 (\textit{top panel}), plus a richness higher than 6 (\textit{middle panel}) or 10 (\textit{bottom panel}). The black points represent the counts, whereas the blue regions show the 68\%  c.i. around the median parametrised composite luminosity functions (cyan lines). The red normalised histograms show the magnitude distributions of the BCGs of all clusters included in each bin. The grey points show the counts when there are fewer than 4.5 clusters contributing, and are not taken into account in the fitting procedure.}
                 \end{center}
                \end{figure}
%_________________________________________________________________________________________

The composite luminosity function including all clusters up to $z=0.67$ is shown in the top panel of Figure \ref{CLF_general}. The black points represent the counts, whereas the blue regions show the 68\%  confidence intervals (c.i.) around the median parametrised composite luminosity function indicated by the cyan line. The red normalised histogram shows the distribution of the BCGs. The grey points show the counts when there are fewer than 4.5 clusters contributing, and are not taken into account in the fitting procedure. The corresponding CLF parameters are presented in the first row of Table \ref{CLF_param_table}.
Within  our magnitude range, we can see that, as expected, the composite luminosity function is well fitted by a single-component Schechter function.

Selecting all clusters with $z<0.67$ includes very poor clusters, and we tested whether this affects the CLF by applying richness cuts at $\lambda_{0.5Mpc}=6$ and  $\lambda_{0.5Mpc}=10$. These limits correspond to the first and second richness bins discussed in the following section. The resulting CLFs are shown in the middle and bottom panels of Figure \ref{CLF_general} and their parameters are presented in Table \ref{CLF_param_table}. 
We can see that when the poorest clusters are discarded, the faint end slope becomes shallower, the characteristic magnitude fainter and the amplitude higher (following the degeneracy between the three parameters). The strong effect on the CLF caused by the poor clusters is driven by the fact that they are up-weighted by the Colless stacking method. Indeed, in Eq. \ref{colless} the individual LFs are weighted by the inverse of their normalisation: $1/\phi_{i0}$.

\begin{figure}
                    \resizebox{\hsize}{!}{\includegraphics{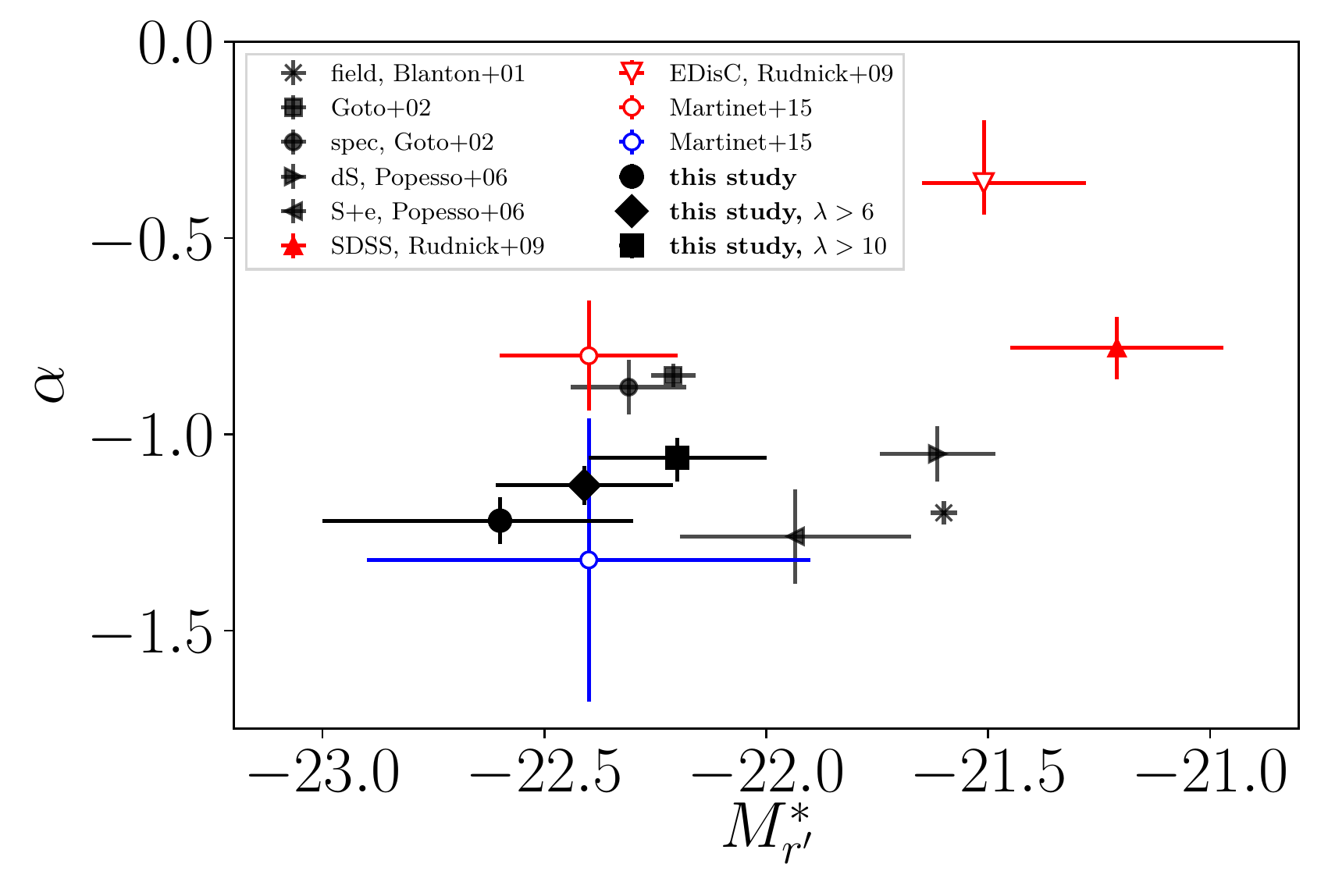}}
                 \caption{\label{alpha_mstar} Comparison of our characteristic magnitude $M^*$ and faint end slope $\alpha$ values obtained for the z<0.67 sample with different richness cuts (black circle, diamond, and square), with those found in the literature. The small cross indicates the field values from \cite{blanton_luminosity_2001}, whereas the other points indicate the values for composite luminosity functions from  \cite{goto_composite_2002},  \cite{popesso_rass-sdss_2006}, \cite{rudnick_rest_frame_2009}, and \cite{martinet_evolution_2015}, including all galaxies (black), only  red sequence galaxies  (red), or only  blue cloud galaxies  (blue). The high redshift samples are indicated by empty markers. Because of the good agreement between the bands used by the different studies, we did not apply any correction. The values are corrected to our cosmology. }
                \end{figure}

In Figure \ref{alpha_mstar}, we compared our parameter values with those found in the literature and presented in Table \ref{tab_biblio}, after correcting to our cosmology. Unfortunately, the $\phi^*$ values are often not mentioned or computed with different units and we thus limited our comparison to the values of $M^*$ and $\alpha$,  even though the three parameters are degenerate. The $M^*$ values from the literature were obtained in different red bands ($R$ from VLT/FORS2 for \cite{martinet_evolution_2015} and $r$ from SDSS for the others), but we checked that the differences in absolute magnitude were small enough that they could be neglected.

We note  that there is a disparity among the values of the  CLF parameters even when limited to the same galaxy population. The origin of the diversity may come from the different cluster samples and/or from the different methods used to construct the CLF. We also have to keep in mind that the parameters are positively correlated, which can explain the tendency to have fainter $M^*$ with shallower $\alpha$.
 We can see that our $M^*$ values are compatible within the errors with the values from  \cite{martinet_evolution_2015} and \cite{goto_composite_2002}, and partially with the value from \cite{popesso_rass-sdss_2006} when fitted with a  Schechter function plus an exponential function (S+e). 
 Our faint end slope values are compatible with the field value from  \cite{blanton_luminosity_2001}, the values from \cite{popesso_rass-sdss_2006}, and the value from \cite {martinet_evolution_2015} found for blue cloud galaxies.
 We note that our faint end slopes are steeper than those obtained with red sequence galaxies. 
Finally, considering the large disparity in the $\alpha$ and $M^*$ values reported in the literature, our values are comparable to those found in previous studies.

 %_________________________________________________________________________________________
         \begin{table*}
        \caption{Schechter parameters $M^*$ and $\alpha$ of field LF and CLFs  from the literature.}             
        \label{tab_biblio}     
        \centering  
                  
        \begin{tabular}{c c c c c c c c c c}      
        \hline\hline             
        Reference & gal. type & radius & method & $N_{clus}$& sample & z & $M^*$ & $\alpha$ \\    
        \hline    \hline                   
           \cite{blanton_luminosity_2001}  & all & field           & spec             &  & SDSS                                        &   z<0.2               &-21.6$\pm$0.03 & -1.20$\pm$0.03 \\
        \hline
           \cite{goto_composite_2002}        & all & 0.75Mpc  & phot      &204  & SDSS CE                       & 0.02<z<0.25          &-22.21$\pm$0.05 & -0.85$\pm$0.03  \\
            \cite{goto_composite_2002}        & all & 0.75Mpc  & spec      & 75 & SDSS CE                        &0.02<z<0.25          & -22.31$\pm$0.13 & -0.88$\pm$0.07     \\
           \cite{popesso_rass-sdss_2006}\tablefootmark{a}$^,$\tablefootmark{b}   & all & r500         & phot      &69  & RASS+SDSS                   &<z>=0.1                &-20.84$\pm$0.13 &  -1.05$\pm$0.07     \\
           \cite{popesso_rass-sdss_2006}\tablefootmark{a}$^,$\tablefootmark{c}   & all & r500         & phot      &69  & RASS+SDSS                   &<z>=0.1                &-21.16$\pm$0.26 &  -1.26$\pm$0.12\\
          \cite{rudnick_rest_frame_2009}   & RS & 0.75Mpc  & colour     &167 & SDSS                        & z<0.06    &-21.21$\pm$0.24 &   -0.78$\pm$0.08   \\ 
          \cite{rudnick_rest_frame_2009}   & RS & 0.75Mpc  & colour     &16  & EDisCS                       & 0.4<z<0.8  &-21.51$^{+0.23}_{-0.14}$ & -0.36$^{+0.16}_{-0.08}$  \\ 
         \cite{martinet_evolution_2015}  & RS & 1Mpc       & photo-z+colour   &16 & DAFT/FADA     & <z>=0.58&-22.4$\pm$0.2 & -0.80$\pm$0.14 \\ 
          \cite{martinet_evolution_2015} & BC & 1Mpc      & photo-z+colour   &6& DAFT/FADA     &<z>=0.62 &-22.4$\pm$0.5  &  -1.32$\pm$0.36  \\ 
        \hline                                   
        \end{tabular}
        \tablefoottext{a}\ These values were obtained using $h_0=1$; we converted them to our cosmology in the figure \ref{alpha_mstar},
        \tablefoottext{b}\ These values correspond to the bright part of a LF fitted using a double Schechter function (dS),
        \tablefoottext{c}\ These values correspond to the bright part of a LF fitted using a  Schechter plus an exponential function (S+e).
        \end{table*}
%_________________________________________________________________________________________

%_________________________________________________________________________________________
%_________________________________________________________________________________________

%##################################################################
%##################################################################
        \subsection{Evolution of the galaxy luminosity distributions with redshift and richness}
        \label{CLF_param_evol}
%_________________________________________________________________________________________
                \begin{figure}[h]
                   \includegraphics[width=88mm]{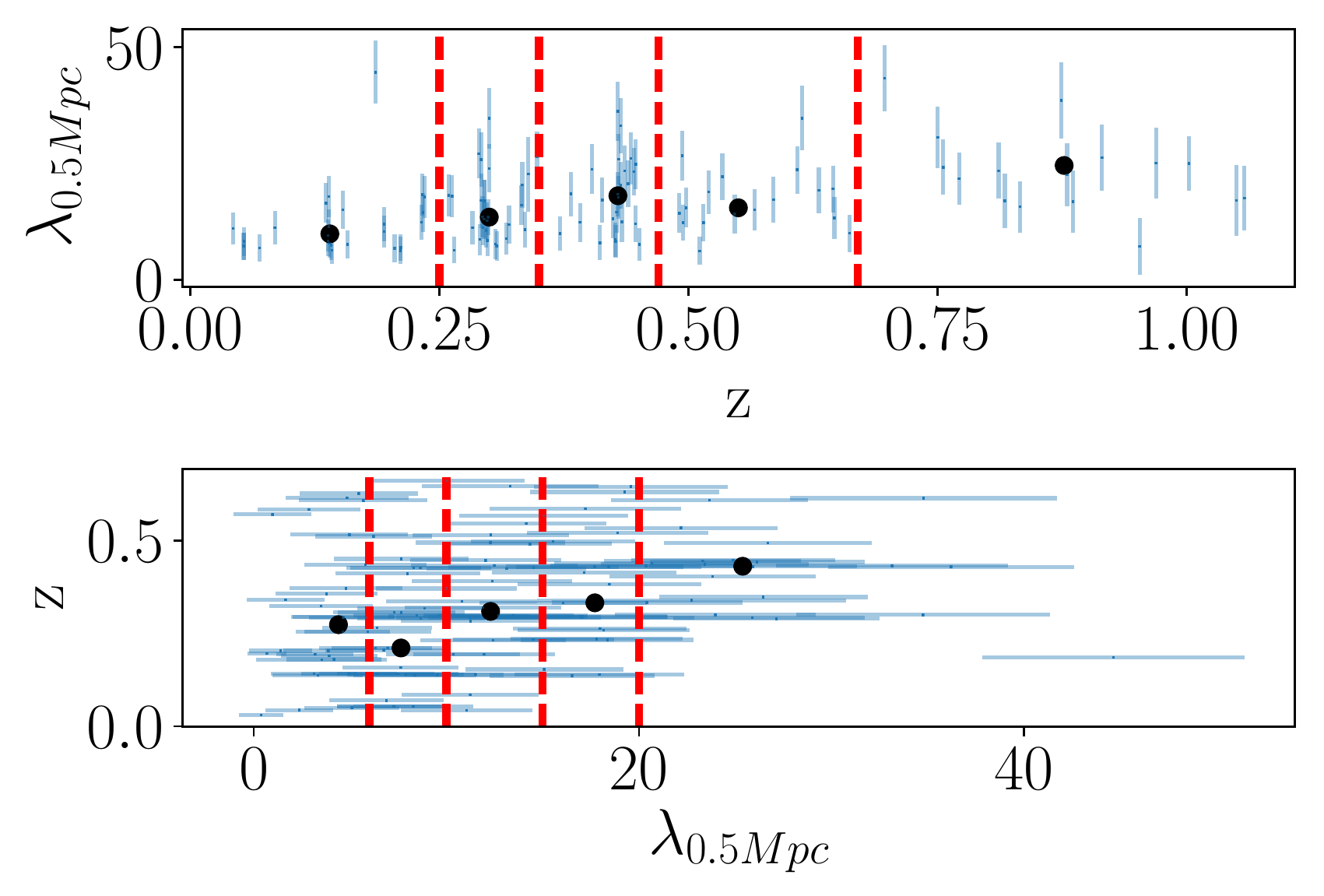}
                 \caption{\label{CLF_bins} Illustration of the CLF bin limits. Measurements for individual clusters are represented by the grey error bars. Bin delimitations are indicated by the red dashed lines.  \textit{Top:} Richness in $0.5$Mpc as a function of redshift, with a minimum richness of $6$. The black dots indicate the median values of the richness in each redshift bin. \textit{Bottom:} Redshift as a function of the richness in $0.5$Mpc, with a maximum redshift of $0.67$. The black dots indicate the median redshift value in each richness bin.}
                \end{figure}
%_________________________________________________________________________________________

%_________________________________________________________________________________________
                \begin{figure*}
                \begin{center}
                  \includegraphics[width=18cm]{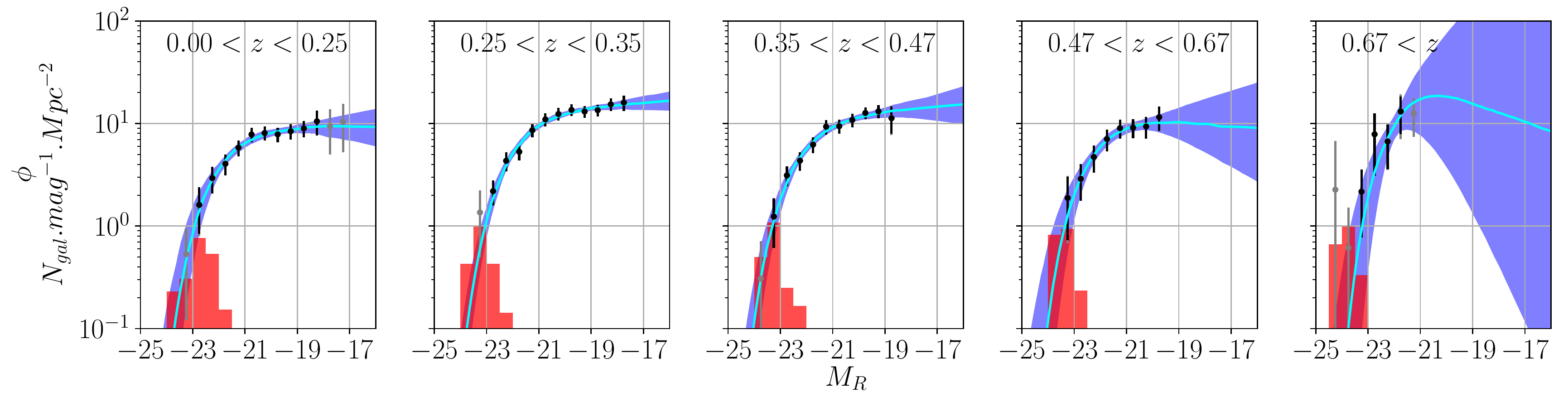}
%                           \put(-69,105){$\lambda>6$}
%                           \put(-165,105){$\lambda>6$}
%                           \put(-261,105){$\lambda>6$}
%                           \put(-361,105){$\lambda>6$}
%                           \put(-456,105){$\lambda>6$}\\
                  \includegraphics[width=18cm]{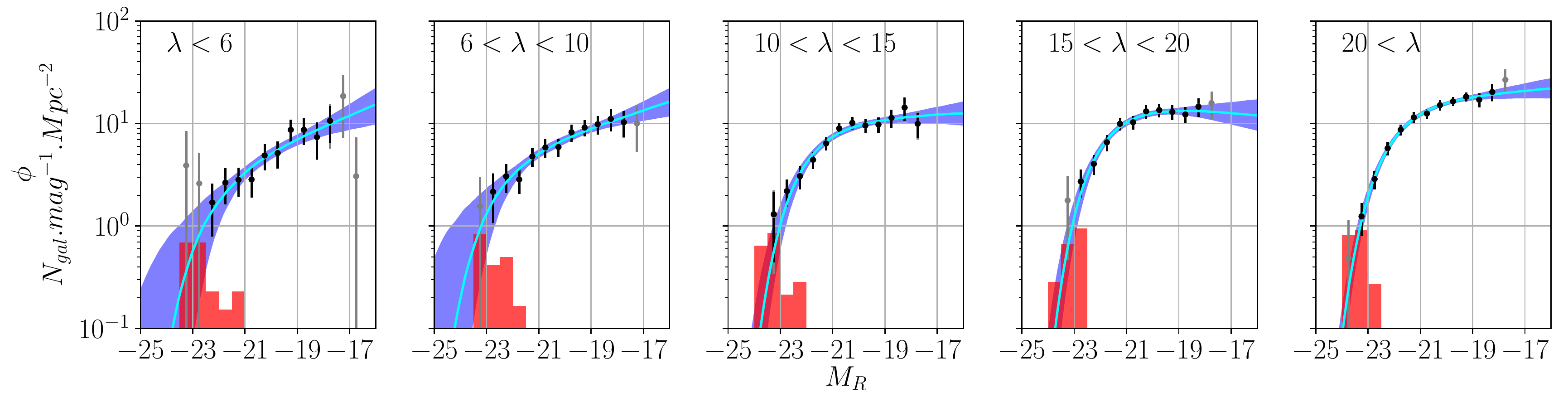}
        %                   \put(-69,105){$z<0.67$}
%                           \put(-165,105){$z<0.67$}
%                           \put(-261,105){$z<0.67$}
%                           \put(-361,105){$z<0.67$}
%                           \put(-456,105){$z<0.67$}
                 \caption{\label{CLF_zbin} Composite cluster luminosity functions in increasing redshift bins, with a minimum richness of $6$ (\textit{top panel}) and increasing richness bins, with a maximum redshift of $0.67$ (\textit{bottom panel}). The black points represent the counts, whereas the blue regions show the 68\%  c.i around the median parametrised composite luminosity functions  (cyan lines). The red normalised histograms show the magnitude distributions of the BCGs of all clusters included in each bin. The grey points show the counts when there are fewer than 4.5 clusters contributing, and are not taken into account in the fitting procedure.}
                 \end{center}
                \end{figure*}
%_________________________________________________________________________________________

%##################################################################
                \subsubsection{Binning choice and parameter evolution fitting procedure}

We studied the evolution of the CLF and BCG distributions with both redshift and richness by binning our cluster sample. Bins in richness were chosen in order to contain roughly the same number of objects, and bins in redshift were defined in order to have the median redshift increasing by approximately  the same amount. The top panel of Figure \ref{CLF_bins} shows the richness as a function of redshift, with bin limits and median richness values in each bin. The opposite is shown in the bottom panel. Further information on the bins can be found in Table \ref{CLF_param_table}. We applied a redshift cut at $z=0.67$ when binning in richness, and a richness cut at $\lambda_{0.5Mpc}=6$ when binning in redshift in order to remove possible contamination by ultra poor or misclassified clusters. However, we found that our results are unchanged, albeit  noisier, if we do not apply the richness cut.

Composite luminosity functions in increasing redshift and richness bins are shown  respectively in the top and bottom panel of Figure \ref{CLF_zbin}. The black points represent the counts, whereas the blue regions show the 68\%  c.i. around the median parametrised composite luminosity function, indicated by the cyan line. The red normalised histogram shows the distribution of the BCGs. The grey points show the counts when there are fewer than 4.5 clusters contributing, and are not taken into account in the fitting procedure. The corresponding CLF parameters are presented in Table \ref{CLF_param_table}.

As we aim to investigate the evolution of the CLF and BCG distribution parameters with redshift and richness separately, we need to consider the steep selection function of our X-ray cluster sample. As can be seen in Figure \ref{CLF_bins}, richness and redshift are indeed linked: we tend to detect richer clusters at high redshift and poorer ones at lower redshift  because of biases affecting X-ray flux limited samples arising from selection and volume effects \citep[see][XXL~Paper~III, for details on selection bias in XXL]{giles_xxl_2015}. 
Therefore, to take into account the biases and distinguish between redshift and richness effects, we fitted the two dependences conjointly. For this purpose we assumed the  evolution model 
        \begin{equation}
        \label{model}
        Y= a\cdot log(1+\widetilde{z}) + b\cdot log(\widetilde{\lambda}_{0.5Mpc}) + c
        ,\end{equation}
where $Y$ is  a parameter of the CLF or BCG distribution computed in a certain bin; $\widetilde{z}$ and $\widetilde{\lambda}_{0.5Mpc}$  the median redshift and richness of the same bin; and $a$, $b$, and $c$ the evolution parameters. In this way  we   hypothesised that the median redshift and richness of a cluster subsample were the key parameters to describe the CLF and BCGs distribution in that subsample. 

In order to constrain the evolution parameters $a$, $b$, and $c$ we combined the values from the redshift and richness bins, and thus fitted 10 data points. We symmetrised the error bars and assumed $\Delta log(\phi^*)=\Delta \phi^*/(\phi^*\cdot ln(10))$, but we did not take into account the bin widths. Finally, we fitted the model of Eq. \ref{model} using the {\tt{Curve\_fit}} function from the {\tt{Scipy.optimize}} {\sc{Python}} library, which uses a Trust Region Reflective algorithm and returns the best fit evolution parameters and their covariance matrix, assuming Gaussian likelihood. 

%##################################################################
                \subsubsection{Evolution of the non-BCGs luminosity distribution with redshift and richness}
                \label{CLF_fit_results}
%_________________________________________________________________________________________
        \begin{figure*}
        \begin{center}
           \includegraphics[width=140mm]{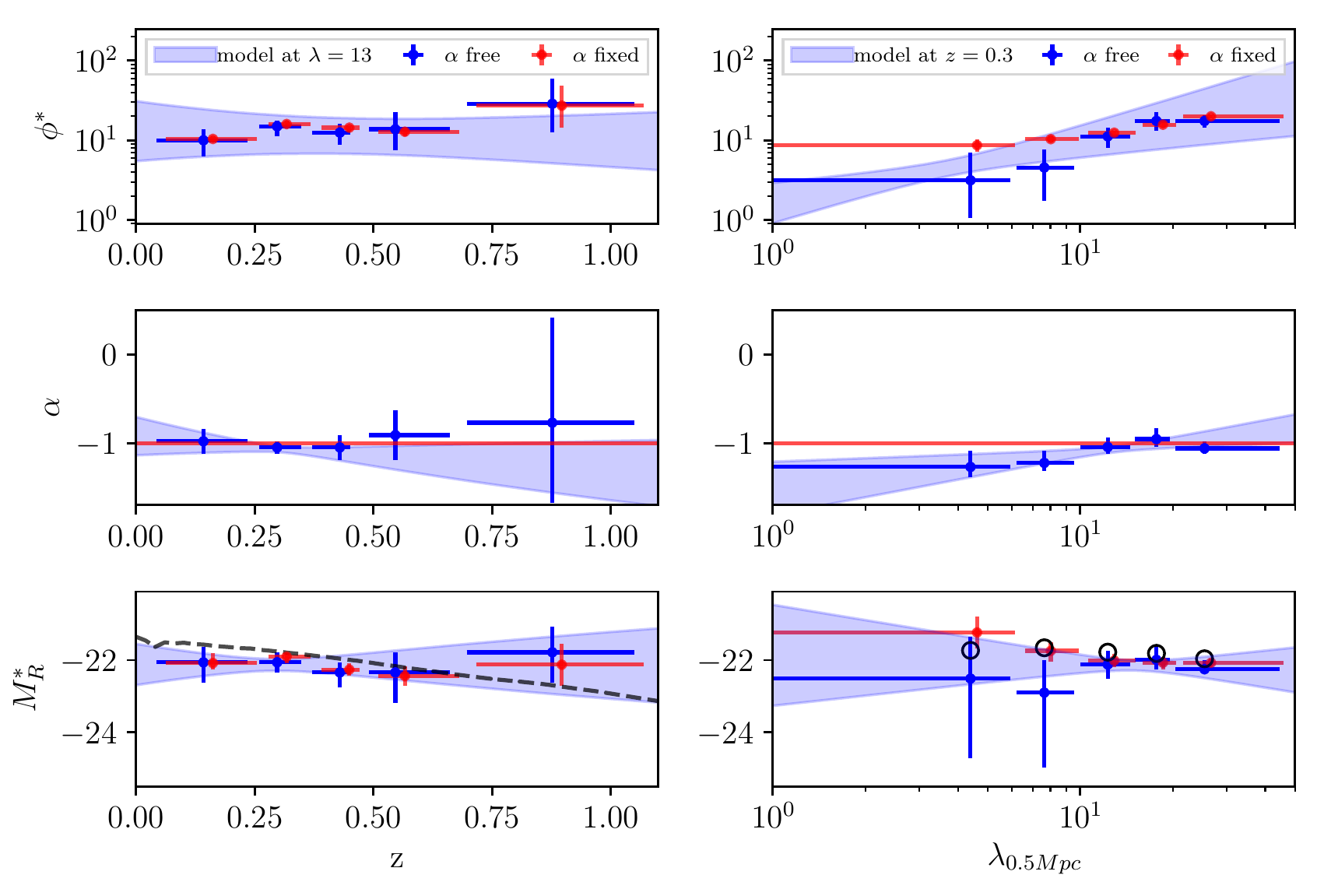}
         \caption{\label{CLF_z_and_rich_param} Parameters of the composite cluster luminosity functions  computed in increasing bins of redshift (left) and richness (right). From top to bottom, the plots show the normalisation $ \phi^*$,  the faint end slope $\alpha$, and the characteristic magnitude $M_R^*$. The vertical error bars indicate the 68\% c.i., whereas the horizontal ones reflect the bin size. The blue (red) points indicate the results when the faint end slope is free (fixed). The shaded blue regions show the evolution models we constrained from Eq. \ref{model} at fixed richness (left) and redshift (right). The dashed black line shows a fiducial model for the evolution of $M*$; the black circles indicate the model values at the median redshifts of the richness bins.}
        \end{center}
        \end{figure*}
%_________________________________________________________________________________________

%---------------------------------------------------------------------------------------------------------------------------------------------------
               \begin{table}
                \caption{Constraints on the evolution of the CLF parameters (see model of Eq. \ref{model}) and associated goodness of fit parameters Q$^{\ref{Q}}$}    
                \label{model_tab1} 
                \begin{tabular}{l || c c c |c   }        
                \hline\hline                
                 &  $a$ & $b$ & $c$ &Q\\
                \hline             
                $log(\phi^*)$ &  $-0.4\pm1.9$ & $0.8\pm 0.4$ & $0.3\pm0.4$ & 0.85\\
                $\alpha $ &    $-1.3\pm1.8$   & $0.4\pm0.3$ & $-1.4\pm0.2$  & $0.87$  \\
                $M_R^* $ &   $-0.1\pm4.9$ & $-0.2\pm1.2$ & $-22\pm1$ & $0.98$      \\                 
                \end{tabular}
                \end{table}
                
%---------------------------------------------------------------------------------------------------------------------------------------------------

We studied the luminosity distribution of the non-BCG cluster members through their composite luminosity functions, shown by the black points and blue shaded regions in Figure \ref{CLF_zbin}.

The  Schechter fit parameters from the  CLFs computed in increasing bins of redshift (left) and richness (right) are shown in Figure \ref{CLF_z_and_rich_param}, where we can see (from top to bottom) the evolution of the amplitude $\phi^*$, the faint end slope $\alpha$, and the characteristic magnitude $M_R^*$. The blue points show the CLF parameters obtained when the faint end slope is set free. 

For each parameter we combined the two data sets and fitted the model from Eq. \ref{model}. The resulting best fit evolution parameters and their $1\sigma$  errors, along with the corresponding goodness of fit parameters Q (probability of obtaining by random chance a $\chi^2$ value equal to or greater than the one we obtained\footnote{\label{Q} $Q$ is defined for a certain $\chi^2$ value as $Q=1-\frac{1}{\Gamma (0.5~n_{dof})} \int_{0}^{\chi^2} t^{0.5~n_{dof}-1}~e^{-t} dt $, with $n_{dof}$ the number of degrees of freedom and $\Gamma$ the gamma function.}) are listed in Table \ref{model_tab1}. 
We represent these evolutionary models by the blue shaded regions in Figure \ref{CLF_z_and_rich_param} by fixing the richness or redshift at the sample median values ($z=0.3$ and $\lambda_{0.5Mpc}=13$). These regions thus show the evolution we would expect if the clusters were all at redshift $z=0.3$ but had different richness (left) or if the clusters all had the same richness $\lambda_{0.5Mpc}=13$ but were at different redshifts (right).

We can see that the amplitude $\phi^*$  increases with richness (at $2\sigma$) and a hint that the faint end slope $\alpha$ becomes shallower with richness (at $1.3\sigma$). Our data are compatible with no redshift evolution for all the CLF parameters, and no richness evolution for the characteristic magnitude $M_R^*$.

Because the faint end slope values are compatible with no redshift evolution and the richness evolution has a low significance, we can fix the value of $\alpha$ to see if we obtain better constraints on the other two parameters, as  is often done in the literature. We thus fixed the faint end slope to a value of $-1$ and repeated the same fitting procedure as before. The $M^*$ and $\phi^*$ values we obtained are shown by the red data points and lines in Figure \ref{CLF_z_and_rich_param} and presented along with their associated goodness of fit parameters in Table \ref{CLF_param_table}.
We can see that the values obtained with the faint end slopes fixed or free to vary are compatible in the redshift bins but not in the richness bins. In the low richness bins the amplitude is higher and the characteristic magnitude fainter when the faint end slope is fixed. This is due to the richness evolution of the faint end slope, which is steeper than $-1$ in these bins. When the faint end slope is fixed, the other two parameters  thus evolve in order to conserve the integrated luminosity. The errors on $\phi^*$ and $M^*_R$ are reduced when $\alpha$ is fixed; however, the comparison of the goodness of fit parameters indicates that setting $\alpha=-1$ is not a good description of the CLF of poor clusters.

We compared our results to the fiducial $M^*$ evolution model used through this study. It is shown by the black dashed line and the black open circles in Figure \ref{CLF_z_and_rich_param}. 
Although a scenario without evolution is not excluded, we can see that the data are compatible on average with the fiducial evolution model with an offset of $\sim$0.5mag (the measured values of $M^*$ being brighter). However, there is a mild  tension at high redshift and at low richness where our values of $M^*$ are respectively too faint and too bright compared to the fiducial model. If statistically meaningful, this would indicate that  the characteristic luminosity of the overall galaxy population in the high redshift and low richness clusters in our sample are not very well represented by the passive evolution of an elliptical galaxy with a burst of star formation at a redshift of 3. We discuss this further in Section \ref{disc_results}.

%##################################################################
                \subsubsection{Evolution of the BCGs luminosity distribution with redshift and richness}
                 \label{BCG_evol}
 %_________________________________________________________________________________________
        \begin{figure}
        \begin{center}
           \includegraphics[width=88mm]{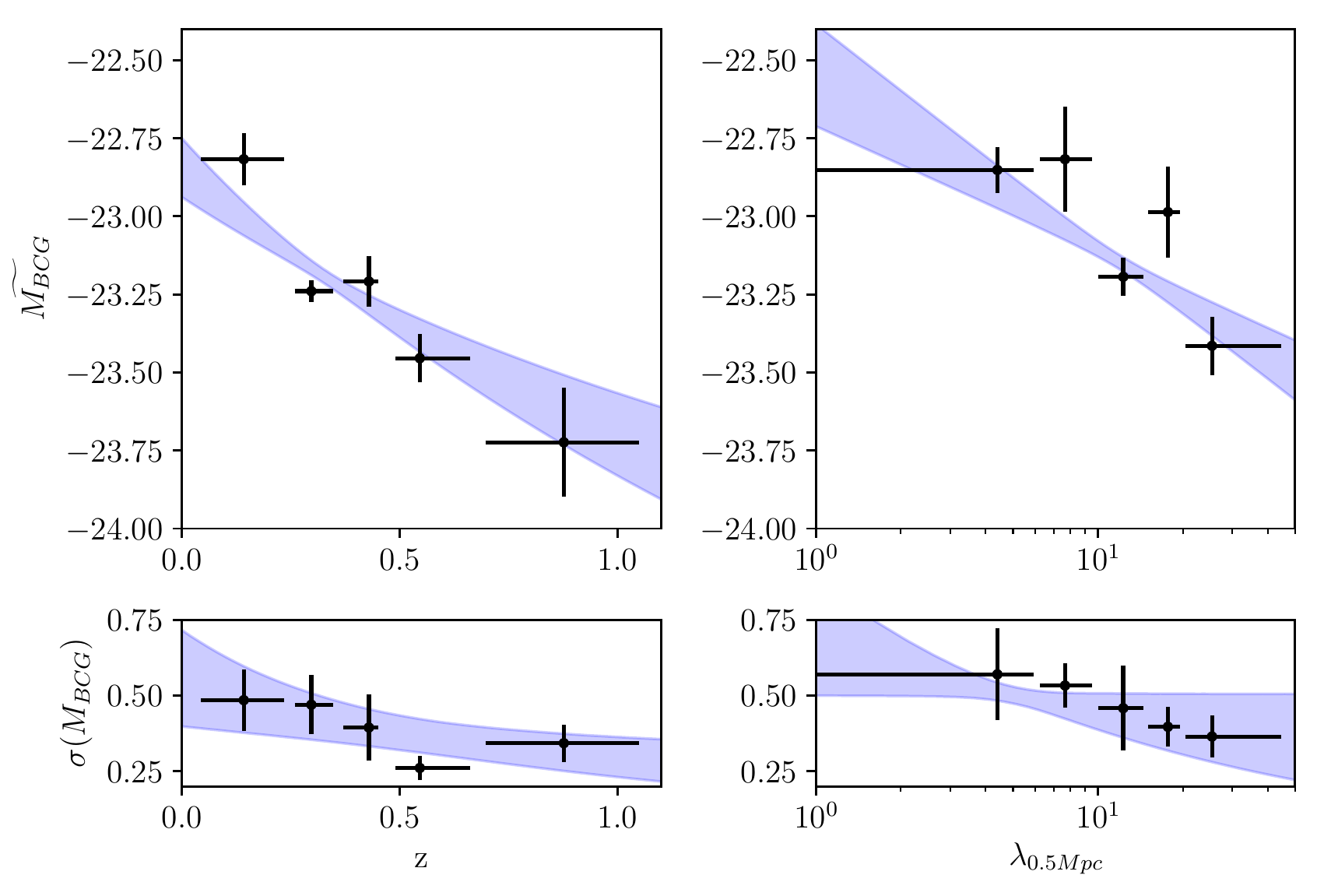}
         \caption{\label{plot_BCG_evol} Parameters of the brightest cluster galaxy (BCG) magnitude distributions, computed in increasing bins of redshift (left) and richness (right). \textit{Top:} Median BCG magnitude.  \textit{Bottom:} Symmetrised scatter of the BCG distributions around the median values. 
         The shaded blue regions show the evolution models we constrained from Eq. \ref{model}, at fixed richness (left) and redshift (right). The vertical error bars indicate the 68\% c.i. obtained from bootstrap, whereas the horizontal error bars reflect the bin sizes.}
        \end{center}
        \end{figure}
%_________________________________________________________________________________________

 The brightest cluster galaxies and the central galaxies in general are known to follow a different distribution compared to the other galaxies and to be better represented by a Gaussian function \citep[see e.g.][]{hansen_measurement_2005,hansen_galaxy_2009, de_filippis_luminosity_2011, wen_dependence_2014}.

 Here we investigate the BCG luminosity distribution in our cluster sample and its evolution with richness and redshift. We first tested the Gaussianity of the distributions and then studied the evolution of  their parameters with richness and redshift.
 
 The distribution of the BCGs in each bin is represented by the red histograms in Figures \ref{CLF_general} and \ref{CLF_zbin}.
 We can see that in some cases the distributions seem quite irregular. We tested the null hypothesis that they follow Gaussian distributions using the D’Agostino and Pearson’s test, based on skew and kurtosis information of the samples. According to this test, the distribution of the BCGs from all the clusters with a redshift $z<0.67$ is very unlikely Gaussian (the p-value is $9e-4$). We found the same conclusion for the poorest clusters: $z<0.67$ \& $\lambda<6$ (p-value=0.09). We concluded that a Gaussian function is not always a good approximation for the BCG distributions in our sample when poor clusters are included. Therefore, we chose to use the median and the 16th and 84th percentiles to describe the distributions rather than the mean and standard deviation.
 
The parameters of the BCGs distributions computed in increasing bins of redshift (left) and richness (right) are shown in Figure \ref{plot_BCG_evol}. The median BCG magnitude is shown in the top panels and the scatter of the BCG magnitude distributions is shown in the bottom panels.  In both cases, the vertical error bars indicate the 68\% c.i. and were computed using bootstrap, whereas the horizontal error bars reflect the bin sizes.
 
To evaluate the evolution of the BCG magnitude distributions and take into account the selection function effects, we again combined the two data sets and fitted the model from Eq. \ref{model}. The resulting best fit evolution parameters and their $1\sigma$  errors, along with their corresponding goodness of fit parameters are listed in Table \ref{model_tab2}. 
We represented these evolutionary models by the blue shaded regions in Figure \ref{plot_BCG_evol} by fixing the richness or redshift at the sample median values ($z=0.3$ and $\lambda_{0.5Mpc}=13$). These regions thus show the evolution we would expect if the clusters were all at redshift $z=0.3$ but had different richness (left) or if the clusters all had the same richness $\lambda_{0.5Mpc}=13$ but were at different redshifts (right).

We can see that our data are compatible with the median BCG magnitude getting brighter with both redshift and richness (at respectively 4 and 3 $\sigma$). There is also a hint that the scatter of the distribution  decreases with redshift (at 1.5 $\sigma$) while being compatible with staying constant with richness. These evolutions are not consistent with a pure passive evolution model. The low and moderate values of the goodness of fit parameters may indicate that the redshift and mass (through richness) are not the only parameters describing the evolution of the BCGs luminosities. This is consistent with the study of \citetalias{lavoie_xxl_2016}, based on the XXL-100-GC sample, where the authors found that the relation between clusters and BCGs masses depends on the clusters' dynamical state.

The scatter of the BCGs magnitude distributions $\sigma (M_{BCG})$ found is $\sim 0.6$ mag for poor clusters and $\sim 0.4$ mag for rich clusters (equivalent to respectively $\sigma$(log L$_{BCG})\sim 0.25$ and $\sigma$(log L$_{BCG})\sim 0.15$). 
\cite{hansen_galaxy_2009} also studied the evolution with richness (and mass) of the BCGs median luminosity and scatter in their low redshift cluster sample. They found that the BCG luminosities increased with the  richness (and mass), while the scatter of the distribution decreased. Their scatter values, $\sigma$(log L$_{BCG})\sim 0.23$ for the poorest clusters and $\sigma$(log L$_{BCG})\sim 0.17$ for the richest clusters, are completely consistent with our findings. \cite{wen_dependence_2014} found a BCG magnitude scatter value of $0.36$ mag in their study of a large sample of rich SDSS clusters, which is again consistent with what we obtained for our richest clusters. 

We conclude that the BCG luminosities is an increasing function of both the redshift and richness, and find a hint that the diversity of BCG luminosity among clusters  decreases predominantly with cluster redshift.

%---------------------------------------------------------------------------------------------------------------------------------------------------
               \begin{table}
                \caption{Constraints on the evolution of the BCGs distributions (see model of Eq. \ref{model}) and associated goodness of fit Q$^{\ref{Q}}$}    
                 \resizebox{0.5\textwidth}{!}{
                \label{model_tab2} 
                \begin{tabular}{l || c c c |c   }        
                \hline\hline                
                 &  $a$ & $b$ & $c$ &Q\\
                \hline           
                $\widetilde{M_{BCG}}$ &  $-2.8\pm0.7$ & $-0.6\pm0.2$ & $-22.2\pm0.1$ & 0.04\\
                $log(\sigma (M_{BCG})) $ &    $-0.9\pm0.6$   & $-0.2\pm0.2$ & $0.0\pm0.2$  & $0.52$  \\             
                \end{tabular}
                }
                \end{table}     
%---------------------------------------------------------------------------------------------------------------------------------------------------

%---------------------------------------------------------------------------------------------------------------------------------------------------
                \begin{table*}
                \caption{Parameters of the composite luminosity functions for cluster selection as stated in the first column. The first block of columns indicates the bin information: number of objects, median redshift, richness, and mass $M_{500,scal}$ (in units of $10^{14} M\odot$, see Section \ref{param} for  definition). The second block indicates the results of the fit of the composite luminosity function: amplitude $\phi^*$ (in units of $N_{gal}\cdot mag^{-1}\cdot Mpc^{-2}$), faint end slope $\alpha$, characteristic magnitude in the $r'$ band $M_R^*$, and goodness of fit parameter $Q$. The third block indicates the results of the fit of the composite luminosity function when $\alpha$ is fixed to $-1$. The values are the median of the marginalised distribution and the errors correspond to 68\% c.i. around the median (see Section \ref{fitting_process}). The goodness of fit parameters$^{\ref{Q}}$ are computed using the minimum $\chi^2$ value in the grid (see Section \ref{fitting_process}).}  
                \label{CLF_param_table} 
                \centering                         
                \begin{tabular}{l||c c c c |c c c c |c c c }        
                \hline\hline          
                  &n& $\widetilde{z}$ & $\widetilde{\lambda}$  & $\widetilde{M_{500,scal}}$ &  $\phi^* $ & $\alpha$ & $M_R^*$ & $Q$ &  $\phi_{\alpha fixed}^* $ &   $M_{R, \alpha fixed}^*$ & $Q_{\alpha fixed}$\\ 
                \hline             
                $z<0.67$             & 121 &  0.30 & 11.9 & 1.00  &  $ 8_{-2 }^{+2 }$ &  $ -1.22 _{-0.06 }^{+0.06 }$  & $ -22.6 _{-0.4}^{+0.3}$ & $0.99$ \\ [0.1cm]
                $z<0.67~\&~\lambda>6$             & 95 &  0.32 & 13.5 & 1.14  &  $ 11_{-2 }^{+2 }$ &  $ -1.13 _{-0.05 }^{+0.05 }$  & $ -22.4  _{-0.2}^{+0.2}$ &$ 0.99$ \\ [0.1cm]
                $z<0.67~\&~\lambda>10$             & 71 &  0.37 & 17.1 & 1.27  &  $ 14_{-2 }^{+2 }$ &  $ -1.06 _{-0.06 }^{+0.05 }$  & $ -22.2 _{-0.2}^{+0.2}$ & $0.97$ \\ [0.1cm]
                \hline
                $\lambda>6~\&~0.00<z<0.25$  & 26   &  0.14 & 9.9 & 0.75 &  $ 10_{-4 }^{+4 }$   &  $ -0.98_{-0.14}^{+0.14}$  & $-22.1_{-0.6}^{+0.4}$ &  $0.99$    & $ 10.4_{-0.8}^{+0.8}$  &  $-22.1_{-0.2}^{+0.3}$ & $0.99$    \\[0.1cm]
                $\lambda>6~\&~0.25<z<0.35$  & 28   &  0.30 & 13.5 & 1.15 &  $ 15_{-4 }^{+3 }$ &  $-1.05_{-0.07}^{+0.07}$ & $-22.1_{-0.3}^{+0.3}$ &   $0.95$    &  $16_{-0.8}^{+1.6}$ &  $-21.9_{-0.2}^{+0.1}$ & $0.97$     \\[0.1cm]
                $\lambda>6~\&~0.35<z<0.47$  & 24   &  0.43 & 18.1 & 1.37 &  $ 13_{-4 }^{+4 }$ &  $-1.05_{-0.14}^{+0.14}$  & $-22.3_{-0.4}^{+0.3}$  &   $0.97$  & $ 14.4_{-1.6}^{+0.8}$ &  $-22.3_{-0.2}^{+0.2}$ & $0.98$    \\[0.1cm]
                $\lambda>6~\&~0.47<z<0.67$  & 17   &  0.55 & 15.5 & 1.14 &  $ 14_{-6}^{+9 }$ &  $-0.91_{-0.28}^{+0.28}$ & $-22.3_{-0.8}^{+0.6}$  &   $0.98$  & $ 12.8_{-1.6}^{+1.6}$ &  $-22.4_{-0.3}^{+0.3}$ &  $0.99$    \\[0.1cm]
                $\lambda>6~\&~0.67<z$             & 12   &  0.88 & 24.6 & 2.55 &  $ 29_{-16}^{+31}$ &  $ -0.77_{-0.91 }^{+1.19}$ &   $-21.8_{-0.8}^{+0.7}$  & $0.36$  &  $ 27.2_{-13}^{+22}$ &   $-22.1_{-0.6}^{+0.6}$ &   $0.60$   \\[0.1cm]
                \hline           
                $z<0.67~\&~\lambda<6$          & 26   &  0.27 & 4.4  & 0.64 &  $ 3_{-2}^{+4}$ &   $-1.27 _{-0.12 }^{+0.18}$  & $ -22.5_{-2.2}^{+1.2}$ & $0.92$ &  $ 9_{-1}^{+2}$      &  $-21.2 _{-0.5}^{+0.5}$ & $0.63$ \\[0.1cm]
                $z<0.67~\&~6<\lambda<10$    & 24   &  0.21 & 7.6 & 0.71 &  $ 5_{-3}^{+3}$ &   $-1.23_{-0.09}^{+0.14}$   &  $-22.9_{-2.1}^{+0.9}$ & $0.98$ &  $ 10_{-1}^{+1}$     &  $-21.7 _{-0.3}^{+0.2}$ & $0.78$ \\[0.1cm]
                $z<0.67~\&~10<\lambda<15$ & 28   &  0.31 & 12.3 & 1.01 &  $ 11_{-3}^{+3}$ & $-1.05_{-0.08}^{+0.11}$ & $ -22.1_{-0.4}^{+0.4}$  & $0.86$ &  $ 12_{-1}^{+0.8}$  &  $-22.0 _{-0.2}^{+0.2}$ & $0.89$ \\[0.1cm]
                $z<0.67~\&~15<\lambda<20$ & 21   &  0.33 & 17.7 & 1.21 &  $ 18_{-4}^{+5}$ & $-0.96_{-0.09}^{+0.12}$ & $-22.0_{-0.3}^{+0.4}$ & $0.96$ &  $ 16_{-1}^{+1}$       &  $-22.1 _{-0.2}^{+0.2}$ & $0.98$ \\[0.1cm]
                $z<0.67~\&~20<\lambda$        & 22   &  0.43 & 25.4 & 2.09 &  $ 18_{-3}^{+3}$ & $-1.06_{-0.06}^{+0.08}$ & $-22.3_{-0.1}^{+0.3}$   & $0.99$ &  $ 20_{-1}^{+1}$     &   $-22.1_{-0.1}^{+0.1}$ & $0.99$ \\[0.1cm]                   
                \end{tabular}
                \end{table*}

%---------------------------------------------------------------------------------------------------------------------------------------------------

%##################################################################
%##################################################################
%Section : SYSTEMATIC EFFECTS
%##################################################################
%##################################################################
\section{Study of the systematics}
\label{syst}

As we have seen, the luminosity function parameters found in the literature  vary from one study to another. There are different plausible explanations for this disparity, and in order to make physical interpretations, we  first need to identify, characterise, and reduce possible systematics. In this section we analyse two main sources of systematic effects affecting the luminosity function measurements, one related to the statistical choice used to obtain discrete LF parameters values and the other  related to the way galaxies are selected. We  discuss their implications further in Section \ref{sys_discuss}.

%##################################################################
%##################################################################
         \subsection{Effects induced by the statistical estimators}
         \label{robust_stats}
%_________________________________________________________________________________________
        \begin{figure*}[!t]
          \centering
                 \includegraphics[width=88mm]{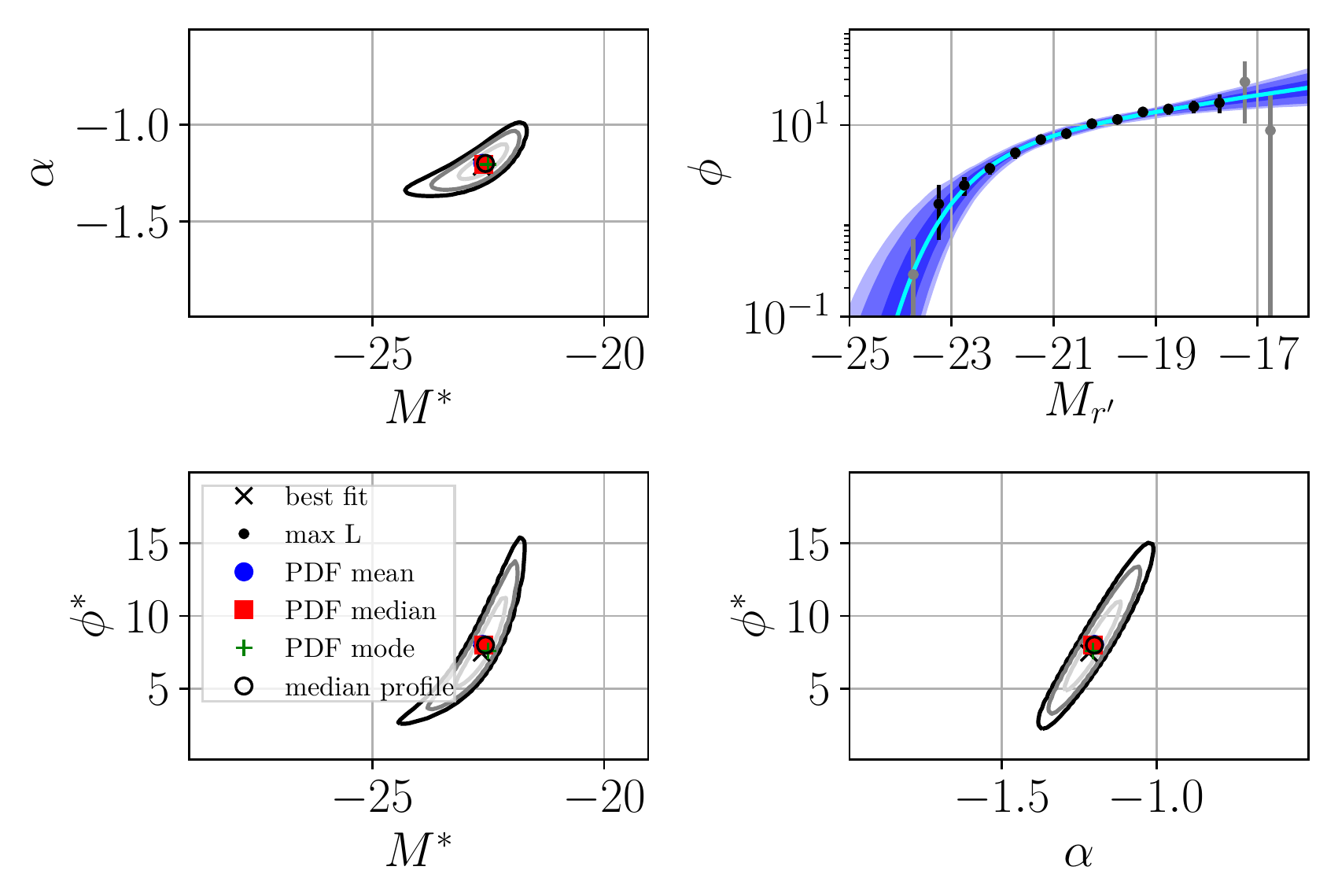}
                 \includegraphics[width=88mm]{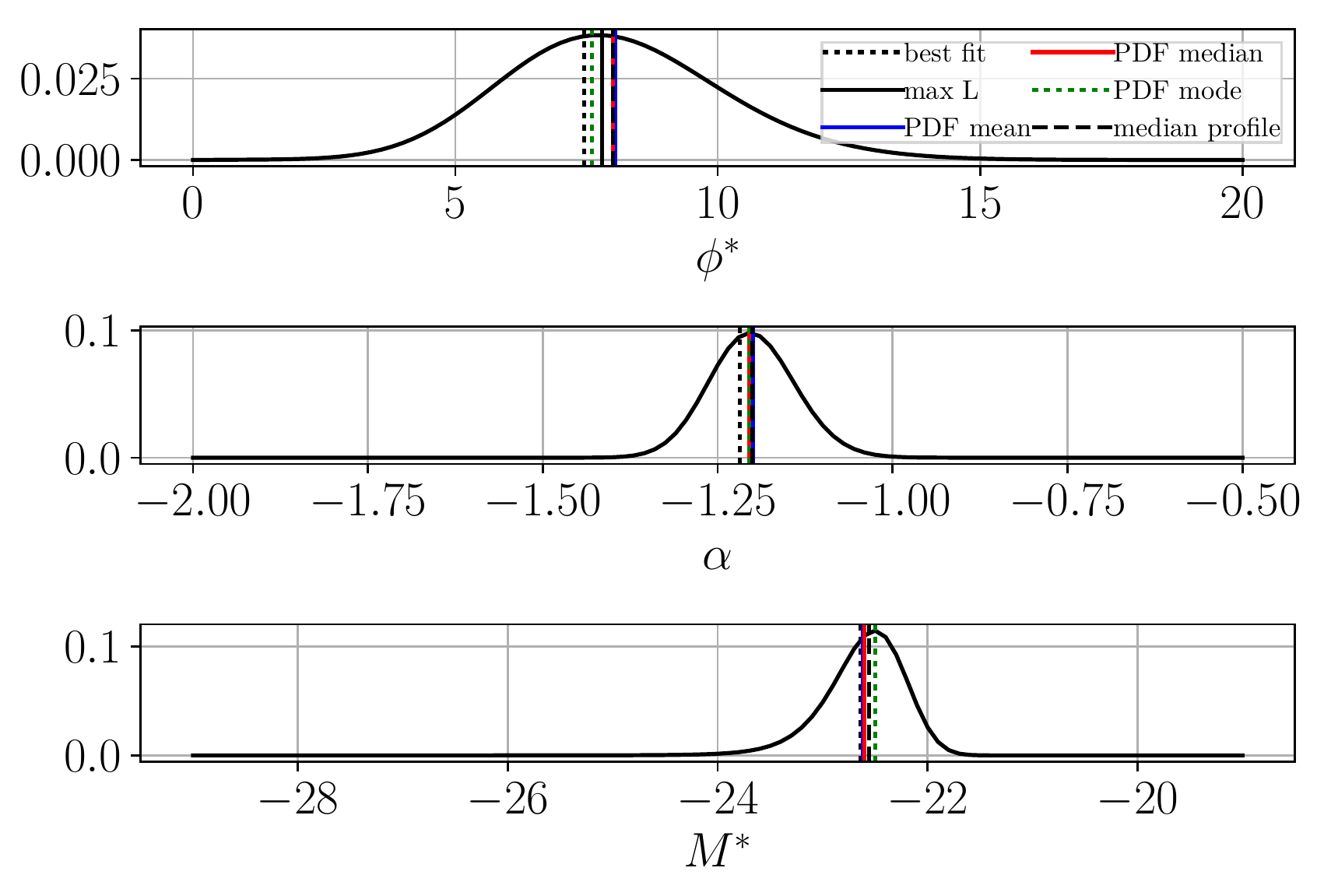}
         
         \caption{\label{general_3para} Illustration of the composite luminosity function fitting procedure for a sample of 121 clusters (general sample in Section \ref{results}).
          \textit{Left panels}: 2D marginalised likelihoods of the Schechter fit parameters, and associated luminosity profile. The contours show the 68\%, 95\%,\ and 99\% levels, and the different statistical values are indicated  in the legend.
           Top left: $(\alpha, M^*)$ marginalised over $\phi^*$; Top right: Posterior CLF shape, the data points are shown in grey and black, the median profile is drawn in cyan, and the blue shaded regions indicate the 68\%, 95\%,\ and 99\% c.i.; Bottom left: $(\phi^*, M^*)$ marginalised on $\alpha$; Bottom right: $(\phi^*,\alpha)$ marginalised on $M^*$.
           \textit{Right panels}: Probability density functions of the Schechter fit parameters after marginalisation. The lines show the different statistical values, as indicated  in the legend. In this sample the different statistical estimators give indistinguishable values.
 }
        \end{figure*}
%_________________________________________________________________________________________

%_________________________________________________________________________________________
        \begin{figure*}[!t]
          \centering
                  \includegraphics[width=88mm]{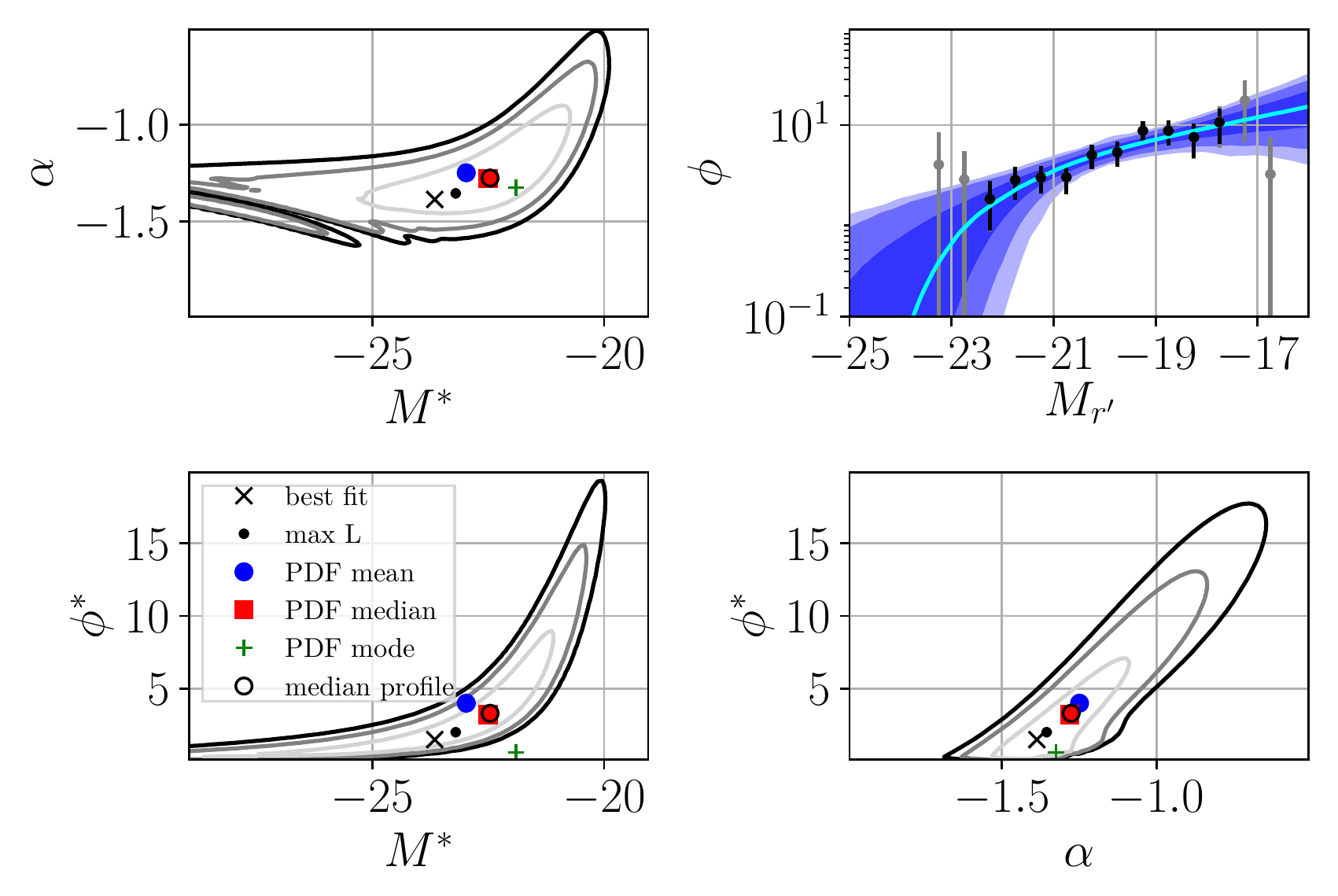}
                  \includegraphics[width=88mm]{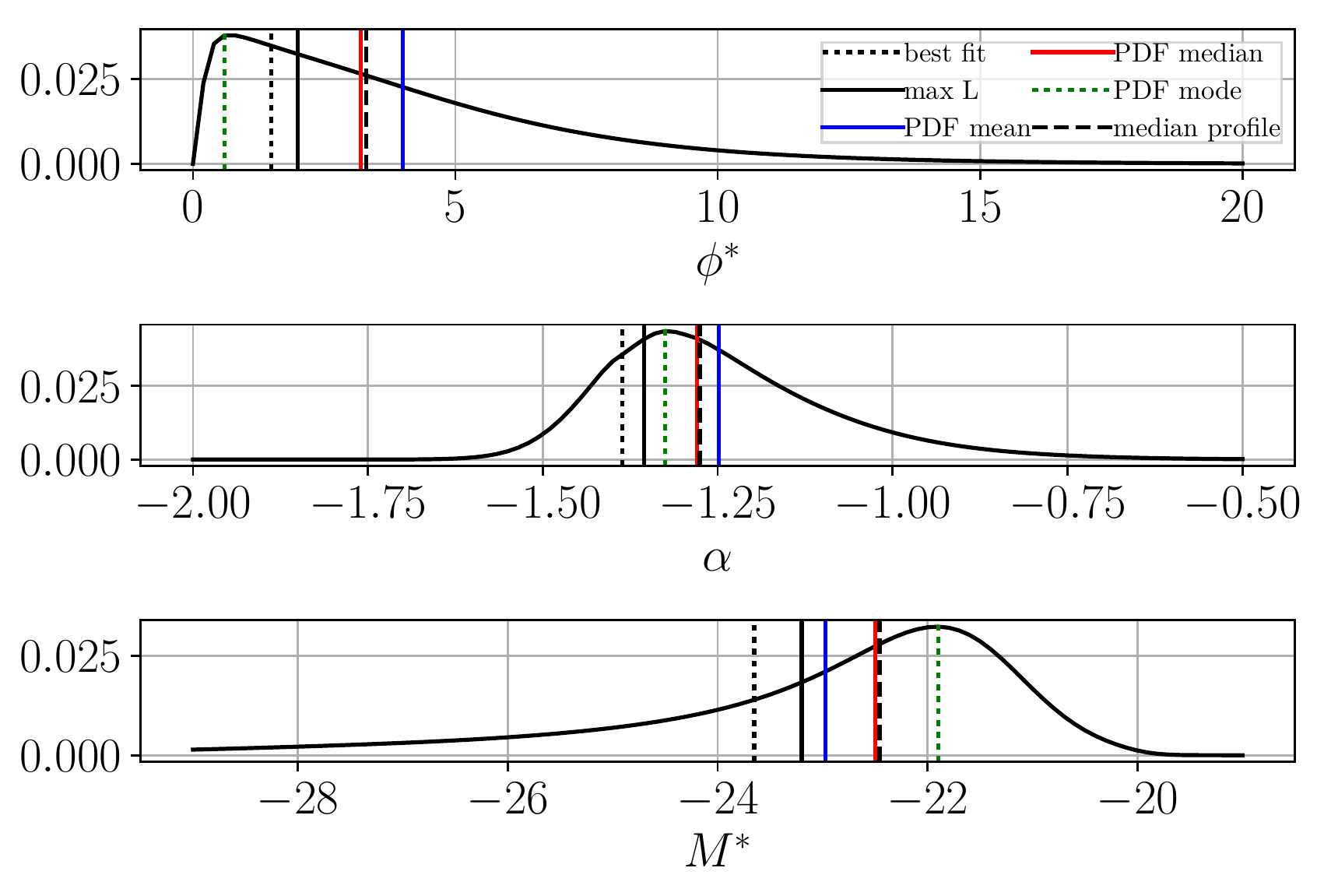}
         \caption{\label{low_rich_3para} Illustration of the composite luminosity function fitting procedure as in Figure \ref{general_3para}, but for a sample of 26 poor clusters (lowest richness sample in Section \ref{results}). In this sample the different statistical estimators give distinguishable values.}
        \end{figure*}
%_________________________________________________________________________________________

We investigated different statistical choices  for extracting discrete parameter values from the likelihood and tested their stability. The statistical properties tested come from the full likelihood (the best fit and the value corresponding to the maximum likelihood), from the PDFs (the mode, median, and mean), and from the Schechter fit of the median luminosity profile.
The best fit value was obtained using the {\tt{Curve\_fit}} function from the {\tt{Scipy.optimize}} {\sc{Python}} library, which uses a Trust Region Reflective algorithm, whereas the value corresponding to the maximum likelihood was computed using $\phi^*-\alpha-M^*$ 3D grids, which is why those two values can differ.

Figure \ref{general_3para} illustrates the composite luminosity function fitting procedure for a sample of 121 clusters (general sample in Section \ref{results}). The  2D marginalised likelihoods of the Schechter fit parameters and the associated luminosity profile are shown in the four left panels, whereas the PDF of the Schechter fit parameters after marginalisation are shown in the three right panels. 
The different statistical values are indicated in the 2D marginalised likelihoods (respectively PDFs) by the following markers: black crosses (dotted black lines) for the best fit; black dots (black lines) for the maximum likelihood; blue points, red circles, and green plus signs (blue, red, and green lines) for the PDF mean, median, and mode; and black circles (dashed lines) for the Schechter fit of the median luminosity profile.

We can see that the shapes of the contours in the 2D likelihood plots can be roughly approximated by ellipses, while the PDFs can be roughly approximated by Gaussian functions. For this sample, the different statistical values are consistent with each other and hardly distinguishable in the plots. 

Figure \ref{low_rich_3para} illustrates again the composite luminosity function fitting procedure, but for a sample of  26 poor clusters (lowest richness sample in Section \ref{results}). For this sample, the statistical errors are larger and the magnitude range a bit smaller than for the larger sample, as can be seen in the luminosity profile plot. The `volume effect' is more pronounced. This causes the contours of the 2D marginalised likelihoods and the PDFs to be much broader, and the approximation by ellipses and Gaussian functions is no longer possible. For this sample, the different statistical values give different values.

This section highlights  that even with the same data it is possible to obtain very different parameter values, albeit  compatible considering the errors, depending on the statistical choice used to obtain discrete values. This is true in particular in the case of a low signal-to-noise sample or a sample with a narrow magnitude range (e.g. at high redshift). 

%##################################################################
%##################################################################
        \subsection{Effects induced by the different galaxy selections}
        \label{robust}

%##################################################################
        \subsubsection{Effects of photometric redshift selection methods}
        \label{robust_method}

 As we show in Section \ref{photometric redshift_calib}, the usual photometric redshift selection methods lead to redshift and magnitude dependent completeness. 
 However,  the impact of these redshift and magnitude dependent completeness values on the LFs shapes is not straightforward because we are not measuring absolute counts, but an excess of galaxies with respect to a background field. 
 In this section we thus investigate the influence of the different photometric redshift selection methods described in Section \ref{photometric redshift_calib} on the shape of the luminosity functions.
 
In order to enhance the signal we stacked the cluster LFs (see Section \ref{method_CLF}) in different redshift bins, to explore possible systematic effects induced by the photometric redshift selections. In order to make a  comparison, we only show redshift bins for which each selection includes the same clusters.

\begin{figure}[h!]
          \includegraphics[width=88mm]{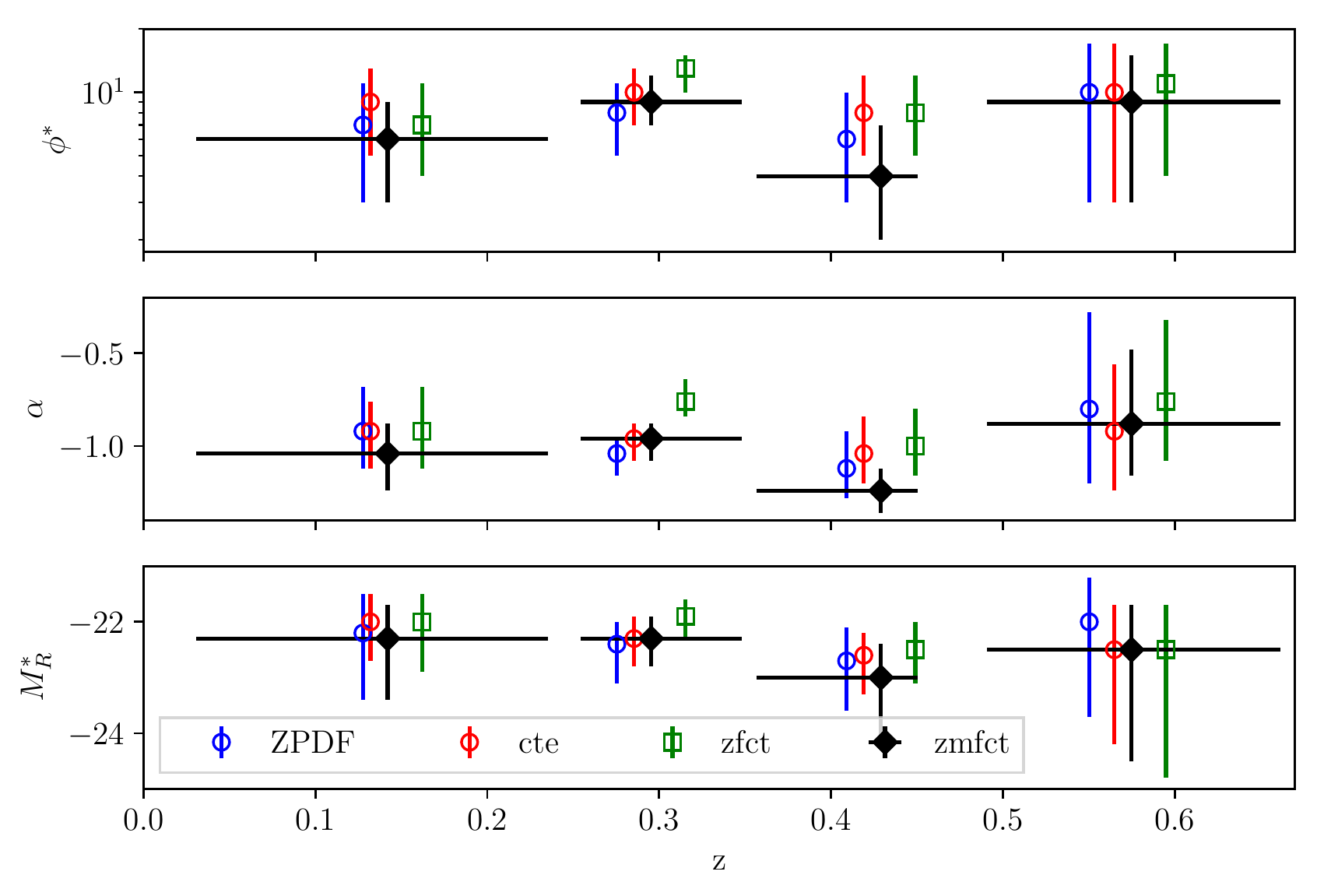}
         \caption{\label{diff_membership_CLF_param} Evolution of the composite cluster luminosity functions parameters with redshift, for different photometric redshift selection methods, as indicated in the legend. The vertical error bars indicate 68\% c.i., whereas the horizontal error bars reflect the bin sizes. The points have been slightly shifted horizontally for clarity. }
        \end{figure}

Figure \ref{diff_membership_CLF_param} shows the parameter evolution as a function of redshift for CLFs constructed using different photometric redshift selection methods:  PDZ errors (\textit{ZPDF}); constant dispersions corresponding to $\sigma_{1/(1+z)}=0.04$ for $i'<22.5$ and $\sigma_{1/(1+z)}=0.08$ for $i'>22.5$ (\textit{cte}); a dispersion computed as a $z$ function (\textit{zfct}); and a dispersion computed as a $(z,i' mag)$ function (\textit{zmfct}).
All the selections are made at the $1\sigma$ or 68\% level. 
From top to bottom, we can see the evolution of the amplitude $\phi^*$, the faint end slope $\alpha$ and the characteristic magnitude $M^*_R$ for the different methods, as indicated in the legend. The vertical error bars indicate 68\% c.i., and the   horizontal ones reflect the bin sizes.

We  note that the CLF profiles and their associated parameters generally agree, considering the error bars. However, as the different selection methods are applied on the same data,  the differences we see between their CLFs are mainly due to systematic errors and not statistical ones.  The relative fraction of the systematic error compared to the statistical error is non-negligible, especially for $\phi^*$ and $\alpha$. In some cases, the systematic error is even higher than the statistical error. 

The differences between the methods are due to the differences in their completeness values as a function of magnitude in each redshift bin. The mean completeness value biases the amplitude $\phi^*$, while the gradient as a function of magnitude biases the faint end slope $\alpha$. The effect is stronger for the \textit{zfct} method, which shows a higher amplitude, shallower faint end slope, and fainter characteristic magnitude with respect to the    \textit{zmfct} method. This is explained by the fact that the \textit{zfct} method shows  the strongest incompleteness gradient between the bright and faint magnitudes at each redshift, as can be seen in the completeness maps of Figure \ref{compl1}.

We conclude that the selection methods having redshift and magnitude dependent completeness can indeed bias the shape of the luminosity function. In our case, the systematic errors due to the selection methods are non-negligible compared to the statistical errors.
%_________________________________________________________________________________________
%_________________________________________________________________________________________

%##################################################################
                \subsubsection{Effects related to the  selection width}
                \label{robust_width}

We defined our selection using photometric redshift dispersions at either 68\%\ or 95\% completeness. If we apply the same dispersion to both the cluster and the background fields, we expect to obtain the same LF shape using one or the other definition, except for the normalisation. Taking a higher dispersion value ensures a higher signal, but may reduce the purity and introduce interlopers. We checked this possible effect by comparing the CLF computed using a dispersion at 68\%\ or 95\% in different richness bins. In order to make a comparison, we only show richness bins for which each selection includes the same clusters.

\begin{figure}
          \includegraphics[width=88mm]{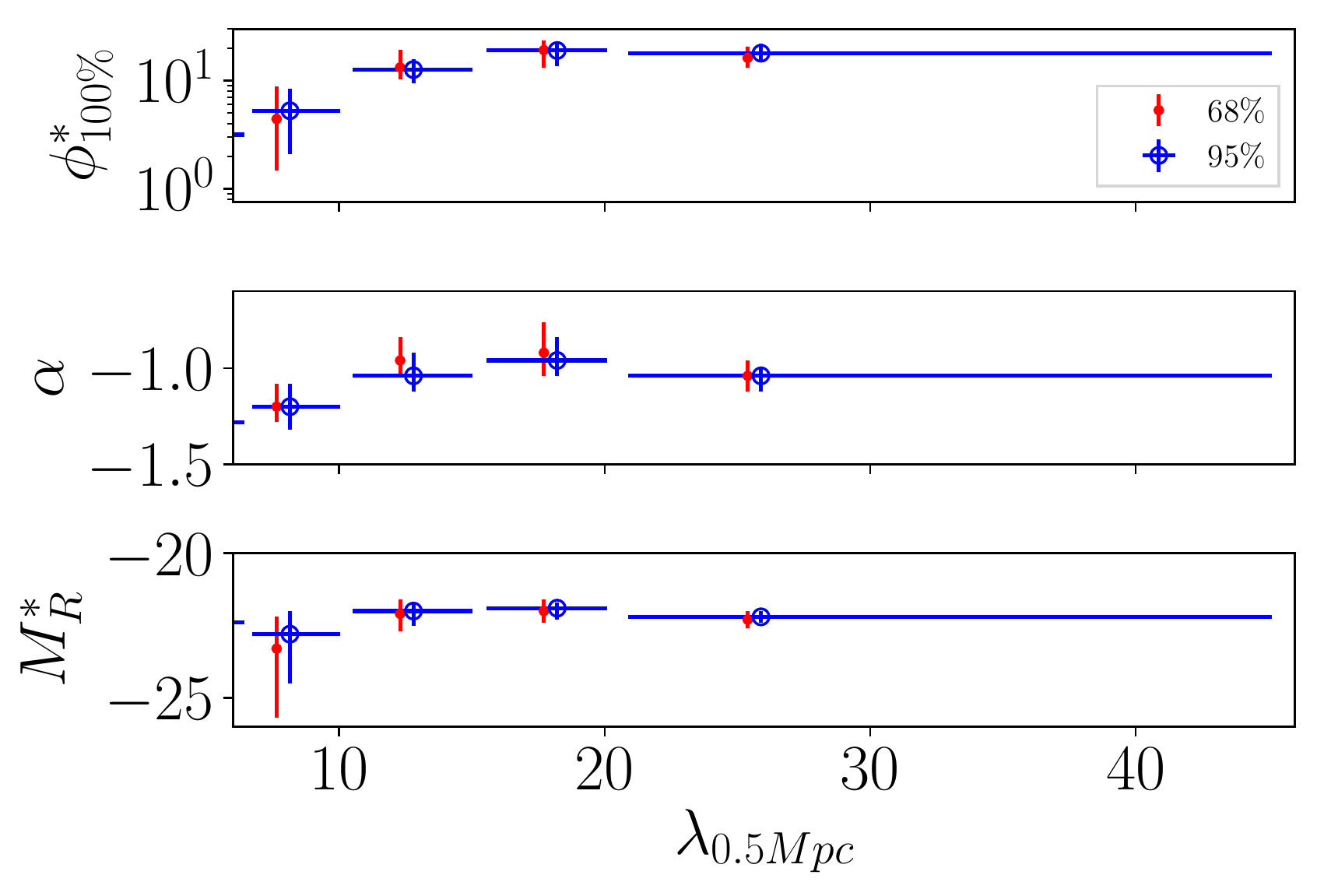}
         \caption{\label{diff_width_CLF_param} Evolution of the composite cluster luminosity functions parameters with richness for two different dispersion widths, as indicated in the legend.  The vertical error bars indicate 68\% c.i., whereas the horizontal ones reflect the bin sizes. The points have been slightly shifted in richness for clarity.  }
        \end{figure}

Figure \ref{diff_width_CLF_param} shows the evolution of the CLF parameters with richness for different photometric redshift selection widths: 68\% in red and 95\% in blue. From top to bottom, we can see the evolution of the amplitude $\phi^*$ normalised to 100\%, the faint end slope $\alpha$, and the characteristic magnitude $M^*_R$. The vertical error bars indicate 68\% c.i., whereas the horizontal error bars reflect the bin sizes.

We note that when we rescale the  amplitude values by the level of completeness we used to compute them, we find that they agree very well ($\phi^*_{100\%} \equiv (1/0.68)\cdot \phi^*_{68\%} \equiv (1/0.95)\cdot \phi^*_{95\%} $), for all the richness bins .
The values of the faint end slope $\alpha$ and characteristic magnitude $M^*_R$ obtained with the two selections are in good agreement. We  also note that the error bars on the parameters are generally larger when computed with the dispersion at $68\%$. 

 Finally, we conclude that the level of completeness ensured by the photometric redshift selection does not affect considerably the shapes of the derived CLFs, except for the amplitude, which   increases proportionally with the completeness.  As the statistical errors are lower with the largest dispersion, we used the dispersion at 95\% to compute the CLFs.
 %_________________________________________________________________________________________
%_________________________________________________________________________________________

%##################################################################
%##################################################################
%Section : DISCUSSIONS
%##################################################################
%##################################################################
\section{Discussions}
\label{disc}

 %##################################################################
%##################################################################
        \subsection{Importance of the systematic effects}
        \label{sys_discuss}
        
%##################################################################
                \subsubsection{Summary of the systematics affecting the CLF measurements}
%---------------------------------------------------------------------------------------------------------------------------------------------------
               \begin{table}
                \caption{Summary of the systematics affecting the faint end slope measurements. The systematics from different origins are estimated differently.}   
                \resizebox{0.5\textwidth}{!}{
                \label{stats_tab} 
                \begin{tabular}{l | c c  }        
                \hline\hline                
                origin of the systematics  & error sys. & error sys./error stat.\\
                \hline             
                estimators stat.\tablefootmark{a}  & $\sigma(\alpha_i) $ & $\sigma(\alpha_i)/\Delta \alpha_{ref}$ \\[0.2cm]
                                                                                                           & $0.07$ & $0.47$ \\
                \hline\hline  
                photo-z selections\tablefootmark{b}  & $<\alpha_i - \alpha_{ref}>_z$ & $<\frac{|\alpha_i - \alpha_{ref}|}{\Delta \alpha_{ref}}>_z$\\[0.2cm]
                $ZPDF$ method & $0.06$ & $0.68$ \\
                $cte$ method   & $0.07$ & $0.61$ \\
                $zmfct$ method & $0.17$ & $1.25$ \\
                \hline\hline  
                selection width\tablefootmark{c} & $<\alpha_{95\%} - \alpha_{68\%}>_\lambda$ & $<\frac{|\alpha_{95\%} - \alpha_{68\%}|}{\Delta \alpha_{68\%}}>_\lambda$  \\[0.2cm]
                                                                                                                  & $-0.03$ & $0.27$ \\
                
                \end{tabular}
                }
                \tablefoottext{a}\ See Section \ref{robust_stats},
                \tablefoottext{b}\ See Sections \ref{photometric redshift_calib} and \ref{robust_method},
                \tablefoottext{c}\ See Section \ref{robust_width}.
                \end{table}
                
%---------------------------------------------------------------------------------------------------------------------------------------------------
In Section \ref{syst} we identified three different origins of systematics affecting the measurements of the luminosity function: the statistical estimators (Section \ref{robust_stats}), the photometric redshift selection methods (Sections \ref{photometric redshift_calib} and \ref{robust_method}), and the width of the photometric redshift slice (Sections \ref{robust_width}). We chose to analyse the  importance of the systematics by estimating the values of the systematic errors and ratios of systematic to statistical error for each origin.
For the sake of conciseness, we only focused on the measurement of the faint end slope $\alpha$, and we present our values in Table \ref{stats_tab}.

The systematic error coming from the different statistical estimators $i$ was estimated using the standard deviation $\sigma(\alpha_i) $ among the different values of the faint end slope $\alpha$ of the CLF containing the poorest clusters. The ratio of systematic to statistical error was estimated by dividing the standard deviation by the symmetrised statistical error of our reference value $\sigma(\alpha_i)/\Delta \alpha_{ref}$, with $\alpha_{ref}$ coming from the median of the PDF. In this case the systematic error value informs us about the spread among the statistical estimators and the ratio of systematic to statistical error tells us about the relative importance of this spread.

In the case of the systematics coming from the different galaxy selections, the differences between the values is not directly due to systematic errors: even if the methods are applied on the same clusters, we do not select exactly the same galaxies (by definition). However, the values are highly correlated and in the following analysis we make the assumption that the differences between them are mainly due to systematics.

The systematic errors coming from the different photometric redshift selection methods were estimated using as reference the $\alpha$ values obtained with our selection method (\emph{zmfct} method) and by averaging over the redshift bins the deviation between the $\alpha$ values from each method $i$ with respect to the references:  $<\alpha_i - \alpha_{ref}>_z$. The ratios of systematic to statistical errors were estimated by dividing each absolute deviation by the symmetrised statistical error of our reference value and averaging over the redshift bins: $<\frac{|\alpha_i - \alpha_{ref}|}{\Delta \alpha_{ref}}>_z$. In this case we see that the systematic error can be null if the values from one method are varying around the reference values and positive or negative in presence of bias, but the systematic error ratio will be null only if there are no differences in the measurements of $\alpha$ induced by one photometric redshift selection method.

Finally, the systematic errors coming from the width of the photometric redshift slice were estimated using as reference the $\alpha$ values obtained with the dispersion at 68\% and averaging the differences with the  values obtained with the dispersion at 95\% over the richness bins:  $<\alpha_{95\%} - \alpha_{68\%}>_\lambda$ and  $<\frac{|\alpha_{95\%} - \alpha_{68\%}|}{\Delta \alpha_{68\%}}>_\lambda$.

From Table \ref{stats_tab} we can conclude that the different statistical estimators give results that have a standard deviation of  $\sigma(\alpha_i)=0.07$, which represents $47\%$ of the statistical errors in the case of a CLF with a low signal-to-noise ratio. The three photometric redshift selection methods lead to faint end slope values that are biased high in average, in particular for the \emph{zmfct} method. The associated averaged systematic error ratios are higher than 60\% and reach 125\% for the $zmfct$ method. Finally, the average difference between the values obtained with the two dispersion widths is low and corresponds to $27\%$ of the statistical error.

%##################################################################
                \subsubsection{Should we care about systematic effects?}

In the previous section we quantified and summarised the systematic effects affecting the faint end slope measurements. We found that the systematic error coming from the different statistical estimators was subdominant but non-negligible in the case of a CLF with a low signal-to-noise ratio. The systematic and statistical error values are expected to decrease strongly with the signal-to-noise ratio and the number of data points. Therefore, the ratio of systematic to statistical error depends on the rate at which these quantities decrease. 

The systematics induced by the different photometric redshift selection methods are biasing high the values of the faint end slope and  
dominate the statistical errors in some cases. They are related to the redshift and magnitude dependent completeness studied at the 68\% level in Section \ref{photometric redshift_calib}. If we increase the width of the selection, e.g. at the 95\% level, the completeness values are closer to $100\%$ and thus the variations are less important. On the one hand, this means that selecting galaxies using a large enough photometric redshift slice  will reduce the systematics coming from the different selection methods (although it introduces other complications, e.g. reducing the purity). On the other hand, the variations of the completeness will be more important when larger magnitude ranges will be probed (e.g. with deeper photometry) and the systematics will dominate the error budget when the statistical errors  decrease (e.g. with a richer or larger cluster sample). Therefore, the systematics induced by the different photometric redshift selection methods need to be taken into account, in particular for studies using  deeper photometry and/or larger cluster sample.

The systematics related to the width of the photometric redshift window were already studied by \cite{crawford_red-sequence_2009}, who found that the faint end slope  becomes steeper when the window  increases (and vice versa), and suggested that this was due either to  the photometric redshift errors being underestimated or to a contamination from field galaxies. However, they used a fixed window and did not take into account the magnitude dependence of the photometric redshift errors. We thus stress that the effect they found is related to their photometric redshift dispersion modelling more than its size and that, as we have shown, the systematics are stronger when the dispersion width is small. 

Finally, if the systematics coming from the width of the photometric redshift slice are negligible for our study, their precise origin has to be investigated in detail if they are no longer subdominant.

%##################################################################
                \subsubsection{Photometric redshift limitations}
                \label{sec_zp_limits}
                
So far we have discussed about the effects of the different photometric redshift selection methods on the shape of the LF. However, these effects are  important not only for LF determination, but also for all the studies that require homogeneous and defined completeness in redshift and magnitude, such as cluster detections, richness estimation, or density profiles construction. 
When appropriate, we can think of checking the photometric redshift quality not as a function of redshift and magnitude, but as a function of other properties such as  galaxy colour, type, or environment.

In the future, larger spectroscopic samples are expected that will allow us to investigate the photometric redshift quality using higher dimensions (e.g. as a combined function of redshift, magnitude, and galaxy type). However, all these analyses require the spectroscopic sample to be representative of the photometric data and thus limit the use of photometric redshift to the redshift-luminosity range covered with spectroscopy. This point also applies to machine learning-based photometric redshift algorithms since the photometric redshift are only representative of the training sample used to derive them.

 In the case of the CFHTLS photometric redshift catalogue, we demonstrated in Section \ref{photometric redshift_calib} that, on average, spectroscopic redshifts were included between the 16th and 84th percentiles of the photometric redshift PDF in less than 68\% of the cases. The confidence intervals coming from the PDF are thus underestimated. Moreover, the photometric redshift PDFs do not reflect the presence of bias between the photometric and spectroscopic redshifts. Therefore, the photometric redshift PDF approach--although very promising because in principle  it allows us to have uncertainties reflecting the signal-to-noise ratio, redshift, and SED of the source and possible multiple peaks--has to be used with caution and improved.

%##################################################################
%##################################################################
        \subsection{Evolution of the CLF parameters}
        \label{disc_CLF}

%##################################################################
                \subsubsection{Are the CLFs representative of the individual LFs?}
                \label{compare_ind}

Throughout this study, we have focused on the evolution of the composite luminosity functions, because of our relatively low mass (and thus low signal-to-noise) cluster sample. However, we show in Section \ref{full_sample} that the stacking method we use (the Colless method) weights poor clusters more. The CLF including all clusters with $z<0.67$ is thus strongly affected by those poor clusters whereas they are not the more numerous. Therefore, we study to what extent the CLFs are representative of the individual LFs, and thus if we can generalise the findings of Section \ref{CLF_fit_results} to the behaviour of individual clusters.

We computed the LF parameters for each cluster in our sample, using the best fit statistical estimators (because it is easier to compute for low signal-to-noise LF, even though biased with respect to the median of the PDF which was used for the CLFs). In each redshift and richness bin we compared the values of $\phi^*$, $\alpha$, and $M^*_R$ coming from the CLF to the mean, median, and weighted mean of the parameters from the individual LF in the same bins. We used as weights the inverse of the squared parameter errors. We computed the error bars using for the mean: the standard deviation; for the median: $1.253$ times the standard deviation; and for the weighted mean: the weighted standard deviation.

We found that the mean, median, and weighted mean values of the faint end slope are compatible with the value from the CLF considering the errors, whereas only the mean and median were compatible with the CLF values for $\phi^*$ and  $M^*_R$. When studying the LF and CLF with the faint end slope value fixed to $-1$, we found that the mean, median, and weighted mean values of  $\phi^*$ and  $M^*_R$ were compatible with the value from the CLF, considering the errors. In both cases,  $M^*_R$ values were systematically brighter (fainter) with respect to the CLF values when using the mean (weighted mean). In general we found that the median values were closer to the parameter values from the CLFs.
We conclude that CLFs are representative of the median of the individual LFs and that the evolutions discussed in Section \ref{CLF_fit_results} can be generalised to the median behaviour of the clusters LFs.

%##################################################################
                \subsubsection{Comparison with previous studies}
                \label{litt_compare}

In this section we compare our finding about the CLF parameters to similar studies from the literature: 
\begin{itemize}

\item \cite{zhang_galaxies_2017} studied the evolution of the red sequence LF parameters with mass and redshift in a sample of 100 X-ray detected clusters using a hierarchical Bayesian method. Their data are compatible with no mass evolution of the faint end slope and characteristic magnitude, and show  a hint that the faint end slope becomes shallower with redshift at a significance level of $\approx1.9\sigma$.

\item \cite{sarron_evolution_2017} studied the CLF evolution with mass and redshift in a large sample of mostly rich optically detected clusters in the CFHTLS-W1 field.

\item \cite{guglielmo_xxl_2017} (XXL~Paper~XXII) studied  the stellar mass function in XXL-N clusters and in the field using a spectrophotometric catalogue. They did not find any significant difference between the shape of the galaxy stellar mass function in the different environments and for galaxies located in clusters of different X-ray luminosities, above their stellar mass completeness limit.

\item \cite{moretti_galaxy_2015} studied the individual LFs of $72$ WINGS of nearby clusters and found that the $M^*$ values (in the bright part of the LF) showed no correlation with mass proxies (using either X-ray luminosities or velocity dispersions).

\item \cite{lan_galaxy_2015} studied the CLF of a large sample of low redshift SDSS clusters, spanning a wide mass range. They found faint end slope values of $\approx -1$, and no evolution of $M^*$ and $\alpha$ with mass inside $R_{200}$. 

 \item \cite{hansen_galaxy_2009} studied the CLF of a large sample of SDSS optically selected clusters (with detections based on the red sequence) in the redshift range $0.1<z<0.3$ and in a mass range comparable to ours. They found, that the CLF computed in $R_{200}$ showed a faint end slope going steeper and a characteristic magnitude getting brighter with richness, while $\phi^*$ (expressed in volume units) was decreasing. The same tendencies were found when only the red galaxies were selected.

\item \cite{alshino_luminosity_2010} studied the LF of 14 C1 clusters from XMM–LSS (that are part of our sample), looking for evolution with redshift and X-ray temperature. They found that, after removing the effects of redshift (correcting for the Malmquist effect), the temperature-stacked LFs did not exhibit any strong evidence of trends with X-ray temperature, while the faint end slope was becoming shallower with increasing redshift. They found faint end slope values much steeper than in our study, but did not constrain the characteristic magnitude of nearly a third of their systems and did not mention the values of the amplitude.
\end{itemize}

Our data are consistent with no richness dependence of the characteristic magnitude, which is consistent with the findings of \cite{moretti_galaxy_2015},  \cite{lan_galaxy_2015}, and  \cite{alshino_luminosity_2010}, but in apparent opposition with \cite{hansen_galaxy_2009}. We found a hint (at $1.15\sigma$.)  of a positive evolution of the faint end slope with richness.
Considering this low significance value, our values  are still compatible with the findings of \cite{lan_galaxy_2015} and \cite{alshino_luminosity_2010}, but again in opposition to \cite{hansen_galaxy_2009}. The discrepancies between \cite{hansen_galaxy_2009} and our study could be explained by the fact that their CLFs are computed in volume units and that the three parameters are degenerated. Another explanation for these differences may be attributed  to the cluster detection, optically red sequence based versus X-ray detected clusters; the first method may  select more evolved, red sequence dominated systems. 

A comparison of our study with \cite{zhang_galaxies_2017} is not possible directly since we are not using the same galaxy population,  but it would suggest that the mild faint end slope dependence on richness we see in our data is driven by an excess of faint blue cloud galaxies in poor clusters. Finally, our results are also in agreement with the study of \cite{guglielmo_xxl_2017} (XXL~Paper~XXII) at least in the massive (bright) part they probe.

Our data are compatible with no redshift evolution of both the characteristic magnitude and the faint end slope. They are also compatible within the error bars with the values of \cite{sarron_evolution_2017} in their lowest mass bin. Our findings are in tension with those of \cite{alshino_luminosity_2010}; however, we stress that since they did not constrain the characteristic magnitude of nearly a third of their systems, and did not mention the values of $\phi^*$, the steep $\alpha$ values they found and their redshift evolution could arise from the degeneracy between the LF parameters. Again, the comparison with \cite{zhang_galaxies_2017} would suggest that faint blue cloud galaxies balance the  increasing deficit of faint red galaxies with redshift.

 %##################################################################
%##################################################################
        \subsection{Implications of our results and perspectives}
        \label{disc_results}

%##################################################################
                \subsubsection{Implications for the use of clusters in cosmology}

The luminosity function is an essential property of galaxies within clusters, in particular in the context of cluster detection. For instance, many cluster finder algorithms (in particular those based on the matched filter technique) use the cluster radial profile and luminosity function to construct their model \citep[see e.g.][]{1996AJ....111..615P,2007A&A...461...81O,2018MNRAS.473.5221B}. A precise and unbiased determination of the luminosity function is therefore mandatory to optimise the cluster detection. Information about the cluster luminosity function can also be used to make prediction about cluster selection functions in optical surveys \citep[see e.g.][in the case of \emph{Euclid}]{2016MNRAS.459.1764S}. In the present paper we have parametrised the evolution of the composite luminosity function parameters with both redshift and richness, in a wide redshift range and for relatively low mass X-ray selected clusters. We have also found that the CLF evolution is a fair representation of the median behaviour of individual cluster LFs. Our study can therefore be used as a reference for analyses requiring knowledge of the optical cluster luminosity function evolution.

The  LF can also be used to derive optical mass proxies such as the cluster richness or optical luminosity. This is done by  integrating the LF to obtain the galaxy number density or the luminosity \citep[as in e.g.][]{lin_near-infrared_2003} and/or by providing a characteristic galaxy luminosity used as a limit \citep[as done in multiple studies, including the present one, using the values of][]{lin_evolution_2006}. Our results indicate an increase in the characteristic galaxy density with richness and no significant LF evolution with redshift. This is compatible with the redshift invariant mass-richness and mass-luminosity relations (at least below $z\sim1$) found by e.g. \cite{lin_evolution_2006} and \cite{2014A&A...568A..23A}. The strong increase in the BCG median luminosity with redshift and richness compared to the CLF evolution may also indicate that the BCG contributes more to the total luminosity budget of the poorest clusters \citepalias[as also found by e.g.][]{ziparo_xxl_2015} and the highest redshift clusters.

We released the catalogue containing the BCGs positions, redshifts, and magnitudes for the 142 clusters in our sample. This is  precious information as the BCG usually resides at the centre of the cluster potential well and is often used as a cluster centre indicator. The location of the BCG with respect to, for instance, the X-ray centroid can thus be used as a cluster dynamical state proxy \citepalias[as done in e.g.][]{lavoie_xxl_2016}.

%##################################################################
                \subsubsection{Implications for galaxy evolution}

The CLF (BCG excluded) in our cluster sample does not significantly evolve with redshift. The characteristic magnitude is still compatible with the passive evolution of an elliptical galaxy with a burst of star formation at a redshift of 3, at least up to $z\sim0.7$. The fact that the measured characteristic magnitude at high redshift is fainter than expected by the model may be due to an enhancement of the star formation in the bright part of the LF, which would make the assumption of passive evolution inadequate. However, the tension is weak and may also be due to the fact that absolute magnitudes are not well constrained by the photometry at these redshifts (see Section \ref{abs_mag}).
The lack of evolution is compatible with a scenario where the bright part of the LF inside $r_{500}$ is already in place at $z\sim1$ and does not significantly evolve afterwards. It is also consistent with the flattening of the cluster red sequence galaxies LF faint end with redshift \citep[suggested by e.g.][]{de_lucia_buildup_2004, de_lucia_build-up_2007, stott_increase_2007, gilbank_red-sequence_2008, lu_recent_2009, rudnick_rest-frame_2009}, which would be compensated by the increase of the faint blue population. In this case, the number of faint galaxies would stay constant while the ratio of red and blue galaxies changes. 

In opposition to what is found for typical member galaxies, a clear evolution is seen in the median luminosity of the BCGs. We compared the median BCG magnitude to the passive evolution model and found an average offset of $\sim1.3$ mag. However, if the passive evolution model (after applying the offset) fits  the measured median BCG magnitudes relatively well, it is excluded by the evolution models we constrained from Eq. \ref{model}. This indicates that the agreement between the BCG luminosity redshift evolution and pure passive evolution found in e.g. \citetalias{lavoie_xxl_2016} is only apparent, and when the selection biases are accounted for (when the richness dependence is fitted conjointly), the measured redshift evolution of the luminosity is weaker. In \citetalias{lavoie_xxl_2016} the authors found that the star formation of the $z<0.5$ BCGs in the XXL-100-GC sample was  comparable to that of similar mass, passive galaxies in the field. Thus, the luminosity evolution weaker than passive that we see in our sample could be due to star formation happening either at $z>0.5$ and/or in clusters that were not part of the XXL-100-GC sample.

We found that the galaxy density at $M^*$  increases with cluster richness, and a hint that the faint end slope is getting shallower. 
Our results thus require a scenario that reduces the number of faint galaxies while increasing the number of bright ones when a cluster grows in mass (gets richer), since the redshift evolution does not play a role. This could be explained by star formation occurring in faint poor cluster galaxies that act to enhance their luminosity. This would lead to a shallower faint end slope, if not enough faint galaxies are accreted, and an increase in intermediate ($\sim M^*$) galaxies. Another scenario that could be responsible for these results is the accretion of substructures with bright galaxies dominated LFs. Since we do not see any evidence of such objects in our sample, it indicates that if they exist such substructures present X-ray emission below the XXL sensitivity.

Our results indicate that the BCG luminosity is increasing with cluster richness, as also found by e.g. \cite{hansen_galaxy_2009}. This is consistent  with the hierarchical formation scenario, according to which BCGs grow by accretion of smaller galaxies and have masses that scale with the cluster total masses \citepalias[see e.g.][and reference therein]{lavoie_xxl_2016}. 

%##################################################################
                \subsubsection{Perspectives}

Several analyses could be done  to push this study forward. Concerning cluster mass proxies, we are currently studying the relation between richness and optical luminosities with X-ray mass proxies in the XXL framework \citep{Ricci_2018_in_prep}. The BCGs detected in the present paper are used to determine the clusters dynamical state and study how it affects the relations between mass proxies.  Another interesting analysis would be to compute the CLF of optically selected clusters (for instance, with the \textsc{WaZP} cluster finder) in the CFHTLS survey, using the exact same methodology as in the present paper, to investigate the difference in terms of galaxy population between optically and X-ray selected clusters. The BCGs could also be used as cluster centres to investigate the impact of X-ray versus optical centring choice.
These analyses will be preparatory to those that will be performed in the near future with the large experiments to come both at X-Ray and optical/NIR wavelengths (e.g. eRosita, \emph{Euclid}, or LSST). 
 
Concerning galaxy evolution in clusters, a natural perspective is to investigate separately the LF of the red and blue galaxy populations. This would give us great insight into the galaxy evolution scenario in dense environments. The link between cluster LF and dynamical state would also allow us to better understand what happens to  cluster galaxies during mergers. Finally, as shown in \cite{Koulouridis_AGN_submm} (XXL~Paper~XXXV), XXL clusters present an enhanced AGN fraction with respect to more massive clusters. Thus, it would be very interesting to study whether the presence of AGNs impact the cluster galaxy luminosities and star formations. 

%##################################################################
%##################################################################
%Section :CONCLUSIONS
%##################################################################
%##################################################################
\section{Conclusions}
\label{ccl}

In this paper, we have studied the optical LFs of a sample of 142 galaxy clusters detected in X-ray by the XXL Survey and having spectroscopically confirmed redshifts. This unique survey has allowed us to study the LF of clusters spanning a wide range of redshifts and X-ray luminosities (and thus masses). 
We constructed LFs using a selection in photometric redshift around the cluster spectroscopic redshift to reduce projection effects  (Section \ref{photometric redshift_calib}).
The width of the photometric redshift selection has been carefully determined to avoid biasing the LF and depends on both the cluster redshift and the galaxy magnitudes. It was defined to obtain a homogeneous completeness in the redshift-magnitude plane. 
The purity was then enhanced by applying a precise background subtraction (Sections \ref{bck_sub} and \ref{abs_mag}).
 We identified BCGs and analysed completeness magnitudes to define the luminosity range for  computing the cluster LFs (Section \ref{sec_lum_range}). 
We then constructed composite luminosity functions (CLFs;  Section \ref{method_CLF}) and defined richnesses (Section \ref{dens}). We parametrised the LFs obtained by a Schechter function and estimated the parameters using likelihood  3D grids in Section \ref{fitting_process}. 
In Section \ref{full_sample}  we presented the general CLF of our sample, investigating the effects of poor clusters and comparing our values to previous studies.
We then studied in Section \ref{CLF_param_evol} the evolution of the galaxy luminosity distributions with redshift and richness, analysing separately the non-BCG and BCG members. We fitted the dependences of the CLFs and BCG distributions parameters with redshift and richness conjointly in order to distinguish between these two effects.
In Sections \ref{syst}  and \ref{sys_discuss} we identified, quantified, and discussed the implications of two main sources of systematic effects affecting the luminosity function measurements. In Section \ref{disc_CLF} we discussed the representativeness of the CLFs with respect to the individual cluster LFs, we compared our results to previous studies, and discussed their impact on cluster cosmology and galaxy evolution.

Our main findings are summarised here, in order of appearance in the text.

\begin{itemize}
\item In Section \ref{photometric redshift_calib}, we  found that the usual method of selecting galaxies using photometric redshifts, defined by using external calibration or by integrating the PDF,  lead to redshift and magnitude dependent completeness. In Section \ref{robust_method}, we  showed that these non-homogeneous completeness causes the resulting LFs shapes, in particular their amplitudes and faint end slopes, to be biased. Our selection in photometric redshift was defined to obtained a homogeneous completeness in the redshift-magnitude plane and allowed us to construct unbiased LFs.

\item In Section \ref{full_sample}, we applied our method to construct CLFs on our cluster sample (for clusters with z<0.67) and found that it was well  fitted by a single-component Schechter function. We studied the impact of poor clusters on this CLF and found that they tend to steepen the faint end slope and brighten the characteristic magnitude because they are up-weighted by the  stacking method we used \citep[adapted from ][]{colless_dynamics_1989}.
 Considering the large scatter among the $\alpha$ and $M^*$ reported in the literature, our values are comparable with those found by previous studies.

\item In Section \ref{CLF_fit_results}, we studied the redshift and richness dependences of the CLF inside $r_{500}$. We found that  the amplitude $\phi^*$  increases with richness (at $2 \sigma$), and that there was a hint that the faint end slope $\alpha$ was getting shallower with richness (at $1.3\sigma$). Our data are compatible with no redshift evolution for all the CLF parameters, and no richness evolution for the characteristic magnitude $M_R^*$.
 We verified in Section \ref{compare_ind} that the CLFs  were representative of the median of the individual LFs, and that our findings could be thus generalised to the median behaviour of the cluster LFs.
This indicates that the bright part of the LF in the inner region of clusters does not depend much on mass or redshift, except for its amplitude, in the redshift-mass range we probe (about $0<z<1$ and $10^{13}$M$_\odot<M_{500}<5\cdot10^{14}$M$_\odot$). 
 We also found a small tension between our data and fiducial evolution model for $M^*$, for the highest redshift and the poorest clusters.

\item In Section \ref{BCG_evol}, we studied the evolution of the BCG distributions with redshift and richness. Our data are compatible with the median BCG magnitude getting brighter with both redshift and richness (at respectively 4 and 3 $\sigma$) and the scatter of the distribution decreasing with redshift (at 1.5 $\sigma$), while staying constant with richness. This means that BCGs are brighter in richer clusters, and that their luminosities decrease with cosmic time, while it seems that their diversity increases. Those results are not consistent with a passive evolution model for the BCG and favour hierarchical formation scenario.

 \item In Sections \ref{robust_stats} and \ref{sys_discuss}, we showed that due to the special shape of the Schechter parameter likelihood, we can obtain different parameters values, and thus we  introduced systematics using different statistical estimators. This is true in particular when the signal-to-noise ratio of the data is low or when the magnitude range probed is small. This effect can be in part responsible for the large variety of values found in the literature. 

\item In Section  \ref{sys_discuss}, we showed that the systematics introduced by the usual galaxy selection methods using photometric redshifts were expected to become even stronger when using deeper photometry. Those systematics may not only affect the LF determination, but also  cluster detections, richness estimations, or density profile construction, for example. We gave some prescriptions about the correct way to use photometric redshifts in Section \ref{sec_zp_limits}.
 
\end{itemize}

%##################################################################
%##################################################################
%######################ACKNOWLEDGEMENTS############################
%##################################################################
%##################################################################
\begin{acknowledgements}
We are thankful to the anonymous referee for the comments that significantly helped to improve the present paper. XXL is an international project based around an XMM Very Large Programme surveying two $25$ deg$^2$ extragalactic fields at a depth of $\sim 5\cdot 10^{-15} erg\cdot cm^{-2}s^{-1}$ in the [0.5--2] keV band for point-like sources. The XXL website is http://irfu.cea.fr/xxl. Multiband information and spectroscopic follow-up of the X-ray sources are obtained through a number of survey programmes, summarised at http://xxlmultiwave.pbworks.com. 
This work was supported by the Programme National Cosmology et Galaxies (PNCG) of CNRS/INSU with INP and IN2P3, co-funded by CEA and CNES.
This study is based on observations obtained with MegaPrime/MegaCam, a joint project of CFHT and CEA/IRFU, at the Canada-France-Hawaii Telescope (CFHT) which is operated by the National Research Council (NRC) of Canada, the Institut National des Science de l'Univers of the Centre National de la Recherche Scientifique (CNRS) of France, and the University of Hawaii. This work is based in part on data products produced at Terapix available at the Canadian Astronomy Data Centre as part of the Canada-France-Hawaii Telescope Legacy Survey, a collaborative project of NRC and CNRS. 
R.A. acknowledges funding from the CNES post-doctoral fellowship programme.
R.A. acknowledges support from Spanish Ministerio de Econom\'ia and Competitividad (MINECO) through grant number AYA2015-66211-C2-2.
The Saclay group acknowledges long-term support from the Centre National d'Etudes Spatiales (CNES).
M.E.R.C. acknowledges support from the German Aerospace Agency (DLR) with funds from the Ministry of Economy and Technology (BMWi) through grant 50 OR 1608.
This research made use of Astropy, a community-developed core Python package for Astronomy \citep{Astropy2013}, in addition to NumPy \citep{VanDerWalt2011}, SciPy \citep{Jones2001}, and Ipython \citep{Perez2007}. Figures were generated using Matplotlib \citep{Hunter2007}. 
\end{acknowledgements}

%##################################################################
%##################################################################
%############################BIBLIO#################################
%##################################################################
%##################################################################

\bibliographystyle{aa} 
\bibliography{biblio_papier_XXL_LF}

%##################################################################
%##################################################################
%##########################APPENDIX#################################
%##################################################################
%##################################################################

%\include{table_bcg}

\end{document}